\documentclass[aps,pra,twocolumn,showpacs,preprintnumbers,amsmath,amssymb,superscriptaddress,longbibliography]{revtex4-1}

\usepackage{mathrsfs}
\usepackage{amsfonts}
\usepackage{amsmath}
\usepackage{amssymb}
\usepackage{natbib}
\usepackage{bbm}
\usepackage{hyperref}
\usepackage{xcolor}

\usepackage{graphicx}

\begin{document}


\title{Deterministic preparation of supersinglets with collective spin projections}


\author{Ebubechukwu O. Ilo-Okeke}
\affiliation{New York University Shanghai, 1555 Century Ave, Pudong, Shanghai 200122, China}  
\affiliation{Department of Physics, School of Physical Sciences, Federal University of Technology, P. M. B. 1526, Owerri 460001, Nigeria}

\author{Yangxu Ji}
\affiliation{State Key Laboratory of Precision Spectroscopy, School of Physical and Material Sciences, East China Normal University, Shanghai 200062, China}
\affiliation{New York University Shanghai, 1555 Century Ave, Pudong, Shanghai 200122, China}

\author{Ping Chen}
\affiliation{State Key Laboratory of Precision Spectroscopy, School of Physical and Material Sciences, East China Normal University, Shanghai 200062, China}
\affiliation{New York University Shanghai, 1555 Century Ave, Pudong, Shanghai 200122, China}

\author{Yuping Mao}
\affiliation{State Key Laboratory of Precision Spectroscopy, School of Physical and Material Sciences, East China Normal University, Shanghai 200062, China}
\affiliation{New York University Shanghai, 1555 Century Ave, Pudong, Shanghai 200122, China}

\author{Manikandan Kondappan}
\affiliation{State Key Laboratory of Precision Spectroscopy, School of Physical and Material Sciences, East China Normal University, Shanghai 200062, China}
\affiliation{New York University Shanghai, 1555 Century Ave, Pudong, Shanghai 200122, China}

\author{Valentin Ivannikov}
\affiliation{New York University Shanghai, 1555 Century Ave, Pudong, Shanghai 200122, China} 
\affiliation{NYU-ECNU Institute of Physics at NYU Shanghai, 3663 Zhongshan Road North, Shanghai 200062, China}

\author{Yanhong Xiao}
\affiliation{State Key Laboratory of Quantum Optics and Quantum Optics Devices, Institute of Laser Spectroscopy, Shanxi University, Taiyuan, Shanxi 030006, China} 
\affiliation{Collaborative research center on Quantum optics and extreme optics, Shanxi University, Taiyuan, Shanxi 030006, China}
\affiliation{Department of Physics, State
Key Laboratory of Surface Physics and Key Laboratory of Micro and Nano Photonic Structures (Ministry of Education), Fudan University, Shanghai 200433,
China}

\author{Tim Byrnes}
\affiliation{State Key Laboratory of Precision Spectroscopy, School of Physical and Material Sciences,
East China Normal University, Shanghai 200062, China}
\affiliation{New York University Shanghai, 1555 Century Ave, Pudong, Shanghai 200122, China}
\affiliation{NYU-ECNU Institute of Physics at NYU Shanghai, 3663 Zhongshan Road North, Shanghai 200062, China}
\affiliation{National Institute of Informatics, 2-1-2 Hitotsubashi, Chiyoda-ku, Tokyo 101-8430, Japan}
\affiliation{Department of Physics, New York University, New York, NY 10003, USA}

\date{\today}

\begin{abstract}
We introduce a procedure to generate supersinglets, the multipartite generalization of angular momentum singlet states.  A supersinglet is defined as a total spin zero state consisting of $ N $ spin-$ j $ particles. They are highly entangled and have zero spin variance in any direction, and as such are potentially useful for quantum metrology.  Our scheme is based on projective measurements that measure the collective spin of the whole spin ensemble.  A local unitary rotation is applied conditionally on the measurement outcome, such as to maximize the probability of obtaining spin zero on the subsequent measurement.  The sequence is repeated in the $ z $- and $ x $-basis until convergence is obtained towards the supersinglet state. Our sequence works regardless of the initial state, and no postselection is required.  Due to the use of strong projective measurements, very fast convergence towards zero spin variance is obtained.  We discuss an example implementation using quantum nondemolition measurements in atomic ensembles, and perform numerical simulations to demonstrate the procedure.
\end{abstract}

\maketitle

\section{Introduction \label{sec:intro}}

Entangled states in many-body systems have been the interest in several areas of modern physics, both from a scientific and application point of view. In condensed matter physics and high energy physics often one is interested in the low-energy states of a many-body Hamiltonian, which is often a highly entangled state \cite{anderson1987,werlang2010,sachdev2011,huber2015}. In quantum information \cite{nielsen2000}, they are the basis for numerous applications, such as quantum simulation~\cite{bloch2012,georgescu2014,afzelius2015,byrnes2007quantum}, quantum sensing~\cite{giovannetti2004,you2017multiparameter}, magnetometry~\cite{urizar-lanz2013,jing2019}, timekeeping~\cite{jozsa2000,ilo-okeke2018}, quantum networks~\cite{perseguers2008,broadfoot2010}, and tests on the foundational principle of quantum mechanics~\cite{bohm1957}. However, the preparation of such correlated quantum states typically requires extremely high degree of control of the individual state of the atoms, which has been the primary experimental challenge for realizing quantum technologies.  Generally, the roadmap towards realizing large-scale quantum systems is to first develop the technology to a high level such that high fidleities are achieved, then use quantum error correction to overcome the remaining errors \cite{lidar2013quantum,devitt2013quantum}.  

An example of such a many-body entangled state is the supersinglet state \cite{cabello2003supersinglets}.  Supersinglets are defined as states of total spin zero, consisting of $ N $ spin-$j$ particles.  Supersinglets have the property that they are invariant under arbitrary total spin rotations.  The variance of the total spin in any basis is also zero, hence they are an example of a state with zero quantum noise, suggesting uses in quantum metrology.  The simplest example of a supersinglet is the $ N = 2 $ spin-$1/2$ case, with wavefunction $ (|0\rangle | 1 \rangle - |1 \rangle | 0 \rangle)/\sqrt{2}  $.  They are typically highly entangled states, and have been proposed for a wide variety of applications such as cryptography \cite{cabello2003supersinglets}, clock synchronization \cite{jozsa2000,ilo-okeke2018}, quantum metrology \cite{toth2010generation}, quantum teleportation \cite{horodecki1999general,bennett1996purification,pyrkov2014full}, quantum computing \cite{byrnes2012,abdelrahman2014coherent}, decoherence free subspaces \cite{toth2010generation} and plays a fundamental role in performing entanglement purification \cite{bennett1996c}.


There have been several demonstrations and proposals to experimentally generate supersinglets.  One approach has been to use quantum nondemolition (QND) measurements of atomic ensembles containing $ N \sim 10^6 $ atoms to produce a macroscopic singlet state of the atoms \cite{toth2010generation,behbood2013,behbood2014,kong2020}.  
Behbood, Mitchell and co-workers used 
stroboscopic QND measurements along different spin axes to generate a collective singlet state of the hyperfine ground states of $^{87} \text{Rb} $ \cite{behbood2014}. This approach has been applied to both cold atoms~\cite{behbood2014} and a hot interacting atomic gas~\cite{kong2020}.  In another work, the same authors incorporated a feedback mechanism between the stroboscopic QND measurements to provide corrections or adjustments to the quantum state of the atoms based on information from the previous measurement outcome ~\cite{behbood2013}. Typically, postselection is used to infer the the presence of macroscopic singlet state at the end of measurements~\cite{behbood2014,kong2020,behbood2013}. These experiments were based upon the theoretical proposal of Ref. \cite{toth2010generation}, where it was shown that QND measurements can produce squeezing towards a supersinglet-like state.  However, even in this theoretical work, it was not shown that a perfect supersinglet state could be attained even in the ideal case, with only squeezing of spin variables being calculated. 
For smaller $ N $, several methods have been proposed to generate supersinglets using approaches such as cavity quantum electrodynamics \cite{cabello2002,jin2005generation,qiang2011alternative,chen2016fast}.  
	

In this paper, we present a scheme that deterministically generates a supersinglet state.  In our scheme, a projective measurement is performed on the collective spin of the particles in the $ z$-basis, recording the total spin projection.  A local rotation is then performed on half the atoms selected at random conditional on the measurement result. This increases the singlet state admixture, thereby building coherence between the quantum states. This process is repeated until the zero spin outcome is obtained.  Then the process is repeated in the $ x $-basis, again until a zero spin outcome is obtained.  Repeating this sequence of measurements allows for convergence to a supersinglet state (see Fig. \ref{figflow} for the sequence).  An important feature of the scheme is that it works for an arbitrary initial state, and no postselection is required.  The key reason for this is that our sequence has the zero spin state as a unique fixed point and the system keeps evolving until it reaches this state.  Our main proposal is generic and can be performed in principle on any physical system.  We show an example implementation based on QND measurements of atomic ensembles.  


This paper is organized as follows. In Sec.~\ref{sec:Dispersiveimaging} we briefly review the physical system and the required operations to realize our procedure.  In this section we also show how QND measurements could be used to realize the projective measurements in an ensemble of atoms.   In Sec.~\ref{sec:Algorithm} we give the detailed procedure used for the deterministic preparation of supersinglets, and the associated mathematical formalism.  In  Sec.~\ref{sec:Examples} we demonstrate that our procedure works by performing several numerical simulations.   Finally the summary and conclusions are presented in Sec.~\ref{sec:Discussion}.

\section{Collective spin measurements and rotations}\label{sec:Dispersiveimaging}

\subsection{The physical system}

We first describe the types of operations that will be required in order to realize the supersinglets.  Consider a collection of \emph{N} particles of spin-\emph{j}. We may construct a basis for the $ (2j+1)^N$ Hilbert space using the vectors
\begin{align}
| m_1, \dots, m_N \rangle =  \bigotimes_{n=1}^N | j, m_n \rangle  .  
\end{align}
Spin operators on the $ n $th and $ m $th particle satisfy commutation relations
\begin{align}
    [J_n^\alpha, J_m^\beta ] = i \delta_{n m} \epsilon_{\alpha \beta \gamma} J_n^\gamma,
\end{align}
where $ \epsilon_{\alpha \beta \gamma} $ is the Levi-Civita antisymmetric tensor and $ \alpha,\beta,\gamma \in \{ x,y,z \}$.  The states on the $ n $th particle are eigenstates of the operator 
\begin{align}
  J_n^z  | j, m_n \rangle = m_n | j, m_n \rangle .
  \label{collectivezeigenstate}
\end{align}
Here $ m_n \in \{-j,\dots, j \}$ is the $z$-projection spin quantum number for the $ n$th particle. 

It is also possible to construct a basis using collective spin states.  Define the collective spin operators
\begin{align}
   J^\alpha = \sum_{n=1}^N J_n^\alpha,    
\end{align}
again for $ \alpha  \in \{ x,y,z \}$.  The $z$-projection collective spin has eigenstates
\begin{align}
J^z | J, d, m \rangle = m | J, d, m \rangle  ,  
\end{align}
and the total spin squared operator (Casimir invariant)
\begin{align}
J^2 = (J^x)^2 + (J^y)^2 + (J^z)^2 ,
\label{casimir}
\end{align}
which has the eigenvalue relation
\begin{align}
 J^2 | J, d, m \rangle = J(J+1) | J, d, m \rangle . 
\end{align}
Here $ d $ is a label for each distinct $ J $ multiplet \cite{schimpf2010}. For example, when adding three $ j = 1/2 $ spins, there are two ways of obtaining $ J = 1/2$.  The number of orthogonal basis elements is the same in the collective picture, such that
\begin{align}
\sum_{J=J_{\text{min}}}^{J_{\text{max}}}
\sum_{d=1}^{D_J}
\sum_{m=-J}^{J} 1 = (2j+1)^N .  
\end{align}
where we denoted the number of distinct $ J $ multiplets as $ D_J $. The minimum and maximum value of the collective spins are $ J_{\min} = (N \mod 2) j $ and $ J_{\max} = N j $.  

Supersinglet states are then defined as states in the zero total spin sector
\begin{align}
|S_{N,d} \rangle = | J = 0, d, m = 0 \rangle .   
\end{align}
Properties and applications of supersinglets may be found in Ref. \cite{cabello2003supersinglets}.

\subsection{Required controls}
\label{sec:required}

We now describe the types of quantum operations that will be required for the procedure that we introduce Sec. \ref{sec:Algorithm}.  

The first capability that we will require is the ability to perform collective spin projections.  Specifically, we assume that it is possible to perform the projective operator
\begin{align}
    P^{z}_m = \sum_{J=J_{\text{min}}}^{J_{\text{max}}}
\sum_{d=1}^{D_J} | J, d, m \rangle \langle J, d, m | , 
    \label{projzmeasurment}
\end{align}
where the superscript $ z $ denotes the basis of the spin states involved in the projector.  
This projection operator can also be written in the individual spin basis
\begin{align}
 P^{z}_m = \sum_{ \sum_{n=1}^N m_n = m} | m_1, \dots, m_N \rangle \langle m_1, \dots, m_N  |  .  
\end{align}

A second capability that we assume is to perform unitary rotations on either part or all of the spins. Define the collective spin operator for a subensemble of the spins as
\begin{align}
   J^\alpha_{\cal S} = \sum_{n\in {\cal S} } J_n^\alpha 
\end{align}
where $ {\cal S} $ specifies which spins are in the subensemble and $ \alpha \in \{ x, y, z \}$.  The unitary rotation for the subensemble is then 
\begin{align}
U^\alpha_{\cal S} (\theta) = e^{-i J^\alpha_{\cal S} \theta} . 
\label{subensemblerotation}
\end{align}
For a unitary rotation on the whole ensemble, our notation is to omit the subensemble label
\begin{align}
U^\alpha (\theta) = e^{-i J^\alpha \theta} . 
\end{align}
Using this unitary operator, we may define projections along other axes.  In particular we will also use projections along the $ x $-axis, defined as
\begin{align}
P_{m}^{x} = U^y (\pi/2) P_{m}^{z} (U^y (\pi/2) )^\dagger .  
\end{align}

\subsection{Example implementation: atomic ensembles}
\label{sec:atomic}

A variety of physical systems could potentially be used to implement the operations that were introduced in the previous section. We now give a specific physical implementation of how the projective operator can be achieved in atomic ensembles with QND measurements. 


We consider an atomic ensemble containing $ N $ atoms, where the hyperfine ground states can be used to store quantum information.  A typical example would be $^{87} \text{Rb}$, where only the ground states levels $ F=1, m_F=-1$ and $ F=2, m_F=1$ are populated \cite{bohi2009,riedel2010}. The ground states have a long coherence time due to the lack of spontaneous emission.  In this case, each atom can be considered a $ j = 1/2 $ spin, and the atomic ensemble as a whole forms a collective spin.  Either hot or cold atomic ensembles can be used, but a Bose-Einstein condensate cannot be used in this case because the atoms should be distinguishable such that a supersinglet is available.  For degenerate atoms, the atoms form only the maximal total spin, and a supersinglet state does not exist.   

A projective measurement of the form (\ref{projzmeasurment}) can be realized using QND measurements.  QND measurements have been used extensively as a means to perform measurements on atomic systems, and have been used as a way of generating entangled states \cite{hammerer2010,pezze2018,takahashi1999,higbie2005,kuzmich2004,meppelink2010,ilo-okeke2014,ilo-okeke2016}. In particular, one of the important applications in atoms is as a method of creating squeezed states~\cite{appel2009,schleier-smith2010,sewell2012,cox2016,hosten2016,vasilakis2015,moller2017} in atomic ensembles.  In this approach, the light acquires an atomic state-dependent phase shift, which is then interfered after passing through the atoms. 
The technique has been used to propose a scheme for entangling two spatially separated atomic condensates~\cite{aristizabal-zuluga2021} and realizing non-Gaussian correlated states such as macroscopic Schr\"odinger cat-states~\cite{ilo-okeke2021} in atomic condensates. For atomic ensembles, information about the state of the atoms is acquired by the light pulse nondestructively, and may be used to further manipulate the state of the atom \emph{in situ}.  

Several works have already provided a theory of QND measurements, here we give a brief description of the formalism.  We use the exact wavefunction approach developed in~\cite{ilo-okeke2016,ilo-okeke2014} to describe dispersive imaging measurements.  The basic idea is to interact the atomic ensemble with an optical coherent state via the QND Hamiltonian \cite{kuzmich2000,ilo-okeke2014} 
\begin{align}
\hat{H} = \hbar g J^z \hat{n} , 
\label{hamqnd}
\end{align}
where $g$ is the atom light coupling frequency,  $\hat{n} = a^\dagger a$ is the photon number operator of the light, and $ a $ is a bosonic annihilation operator for the photons.  Such an interaction entangles the light and the atoms.  Performing a measurement on the light in a suitable basis, this collapses the wavefunction such that an indirect meausurement of the atoms is made.  

Specifically, consider an optical coherent state $\lvert\gamma\rangle$  that interacts with some arbitrary initial state of the atoms 
\begin{align}
\lvert \Psi_0 \rangle = \sum_{J=J_{\text{min}}}^{J_{\text{max}}}
\sum_{d=1}^{D_J}
\sum_{m=-J}^{J}  \psi_{Jdm}\lvert J,d,m\rangle
\label{initstate}
\end{align}
according to the Hamiltonian (\ref{hamqnd}) for a time $ t $.  If the light then interferes with another coherent state of light $\lvert\chi\rangle$ via a beamsplitter and photon detection is performed, then the final unnormalized state can be calculated to be after a Gaussian approximation \cite{ilo-okeke2016,ilo-okeke2021}
\begin{equation}
		\label{eq:MeasuredState}
		\begin{split}
\lvert \Psi_{n_c,n_d} \rangle  & = \frac{e^{\frac{n_c + n_d - \lvert\gamma\rvert^2 - \lvert \chi\rvert^2}{2}}}{\left(4\pi^2 n_c n_d \right)^{\frac{1}{4}}} \left({ \frac{\lvert\gamma\rvert^2 + \lvert\chi\rvert^2}{n_c + n_d}} \right)^{\frac{n_c + n_d}{2}} \\ 
			&\times e^{-i\frac{\pi}{2} n_d}  \sum_{J=J_{\text{min}}}^{J_{\text{max}}}
\sum_{d=1}^{D_J} \sum_{m=-J}^{J} e^{i(n_c + n_d)(\phi - gt m)}\\
			& \times  e^{i(n_c \phi_{c} + n_d\phi_{d})}  e^{-\frac{1}{2\sigma^2}\left(m - m_0\right)^2} \psi_{J d m} \lvert J,d,m\rangle , 
		\end{split}
	\end{equation}
where $n_c$ and $n_d$ are the photon counts for the two modes after interference of the optical modes.  The probability of this outcome is given by  $P_{n_c,n_d}  = \langle \Psi_{n_c,n_d} \lvert \Psi_{n_c,n_d} \rangle $  is the probability of obtaining $n_c$ and $n_d$ photons in the measurement. Here we defined  phases are defined as 
\begin{equation}
		\label{eq:phases}
		\begin{split}
\phi  & = \frac{gt m}{2} + \frac{\arg(\chi) - \arg(\gamma) }{2} + \frac{\pi}{4} + \frac{\phi_{p}}{2},\\
\phi_{c} & = \arctan\left(\tan\eta \tan\phi\right),\\
\phi_{d}  & = \arctan\left(\frac{\tan\phi }{\tan\eta} \right) . 
		\end{split}
\end{equation}
The phase $\phi_p$ is a phase offset.  We defined $\eta$ as the ratio of the amplitude of the light coherent state,
\begin{equation}
		\label{eq:lightamplitude}
		\tan\eta = \frac{\lvert\chi\rvert - \lvert\gamma\rvert}{\lvert\chi\rvert + \lvert\gamma\rvert}.
\end{equation}

The  Gaussian factor in (\ref{eq:MeasuredState}) makes it clear that the effect of the QND measurements is to modify the amplitudes of the coefficients such that it is concentrated near the maximum value 
\begin{equation}
\label{eq:peakposition}
m_0 = \frac{1}{gt}\left[\arcsin\left(\frac{1}{\cos2\eta} \frac{n_d - n_c}{n_d + n_c}\right) - \arg(\chi) + \arg(\gamma) - \phi_{p}\right].
\end{equation}
The variance of the Gaussian is meanwhile
\begin{equation}
		\label{eq:width}
		\sigma^2 = \left( \frac{g^2t^2}{8} \frac{n_c + n_d}{n_cn_d}\left[(n_c + n_d)^2\cos^22\eta - (n_c-n_d)^2\right]\right)^{-1} .
\end{equation}
The variance of the Gaussian determines whether the measurement that is made is either a strong projective measurement or a weak measurement. If $\sigma$ is large, then the original wavefunction (\ref{initstate}) is only slightly modulated, and produces number state squeezing.  In the strong measurement regime, $ \sigma $ is small, with only a few states having non-zero amplitude in the vicinity of $ m = m_0 $. In the extreme case,  characterized by $\lvert \chi\rvert,\, \lvert \gamma\rvert \gg 1$, only one of the probability amplitudes near $m_0$  is non-zero \cite{ilo-okeke2022}.  This limit of QND measurements acting as a strong projective measurement
can be used to realize (\ref{projzmeasurment}).  

The unitary rotations on half the atoms in the ensemble can be realized using optical Raman pulses illuminating half the spatial region of the atomic ensemble.  In order to only rotate
half the atoms in the ensemble, the operation should complete before the atoms move out of the spatial region that illuminates the ensemble. For a hot atomic ensemble in a cell of size $\sim  1$ cm, the velocities of the atoms are typically $\sim  300 $ m/s. This gives a time window of $\sim  17 $ us, easily accessible using current techniques of Raman pulses. Furthermore, composite pulses \cite{genov2014correction} could be used to enhance the fidelity of the rotations in presence of various inhomogeneities in the atomic ensemble.


\section{Procedure for deterministic preparation of supersinglets}\label{sec:Algorithm}
We now describe the procedure using the projective measurements and unitary rotations of Sec.~\ref{sec:required} to prepare the supersinglet states.

\subsection{The basic idea}

Given an arbitrary initial state $ \rho_0 $, we would like to have a procedure that deterministcally gives us the supersinglet state $ | S_{N,d} \rangle $, using only the collective operations as given in Sec. \ref{sec:required}.   Depending upon the particle number $ N $, there may be more than one distinct supersinglet state. For our purposes, any superposition or mixture of such supersinglet states will suffice.  The important aspect will be that the total spin $ J $ is zero.  We note that in order to have a supersinglet state, we require $ N $ to be an even number.  For the procedure that we introduce here, we will assume $ N $ is even.  Later we examine the effect of odd $ N $ in Sec. \ref{sec:oddncase}.  

The key insight that yields our procedure is the fact that a supersinglet state is invariant under rotations.  That is, 
\begin{align}
e^{-i (\theta_x J^x + \theta_y J^y+ \theta_z J^z)} | S_{N,d} \rangle
 = | S_{N,d} \rangle ,
 \label{rotationinvar}
\end{align}
up to a global phase and $\theta_\alpha $ are arbitrary coefficients.  This means that it is an eigenstate of both projection operators $ P_0^z$ and $ P_0^x$, such that 
\begin{align}
P_0^z | S_{N,d}\rangle &  = | S_{N,d} \rangle \nonumber \\
P_0^x | S_{N,d} \rangle &  = | S_{N,d} \rangle
\end{align}
It follows that if one were to start in the singlet state, the sequence of $ 2M $ projections made alternately in the $x $ and $ z $ basis would be invariant
\begin{align}
(P_0^x P_0^z)^M | S_{N,d} \rangle = | S_{N,d} \rangle .  
\label{singletsequence}
\end{align}
Regarded as a measurement sequence, the probability of this outcome is 1, since the state does not lose any amplitude after the projection.  This invariance under projections in different bases is one of the key ingredients of our procedure. Other states with $ J \ne 0 $ become rotated with each change of basis and does not have the invariance property of (\ref{singletsequence}).  

\begin{figure}[t]
		\includegraphics[width=\columnwidth]{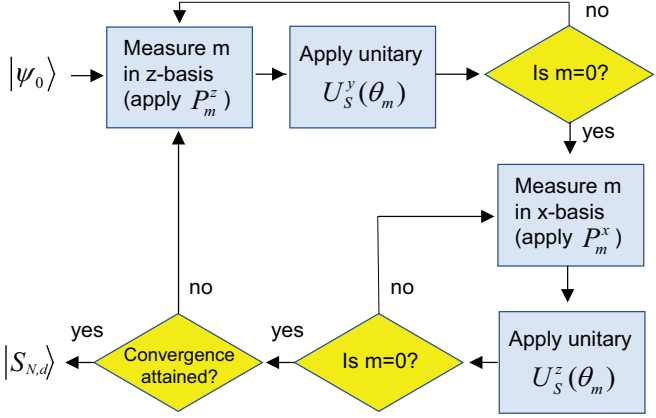}
\caption{The procedure considered in this paper to generate supersinglets.  Starting from an arbitrary initial state $ |\psi_0 \rangle $, the repeat-until-success projection sequence (\ref{tryuntilproj}) is first applied in the $ z $ basis.  Once an $ m = 0 $ is obtained, a similar sequence is performed in the $ x $ basis (\ref{tryuntilsuccessx}).  These two sequences are repeated until convergence is attained.  Convergence is defined as repeated $ m = 0 $ outcomes for both the $ z $ and $ x $ basis projections. }
\label{figflow}
\end{figure}

Starting from an arbitrary state $ | \psi \rangle $, performing a projection in the $ z $-basis gives a probability of getting an outcome  $m $ according to
\begin{align}
p_m = \langle \psi | P_m^z | \psi \rangle . 
\end{align}
Of course in  general there is no guarantee that we will obtain the $ m = 0 $ outcome as in (\ref{singletsequence}).  To overcome this, we perform a conditional unitary rotation after each projection, such as to maximize the probability of getting the desired $ m = 0 $ outcome \cite{toth2010generation}.  Once the $ m = 0 $ outcome is obtained, the next set of projections in the $ x $ basis is performed, where the procedure is repeated, until the $ m = 0 $ outcome is obtained.   Repeating this procedure creates a sequence such that a singlet state is deterministically produced (Fig. \ref{figflow}).

\subsection{The procedure}

We now rephrase the procedure introduced in the previous section mathematically to make it more precise.  We first define the repeat-until-success projection sequence as 
\begin{align}
    {\cal P}^z_{\vec{m}} &  = \prod_{l=1}^L [U^y_{\cal S} (\theta_{m_l}) P_{m_l}^z ] \nonumber \\
  & =  P_0^z U^y_{\cal S} (\theta_{m_{L-1}}) P_{m_{L-1}}^z \dots U^y_{\cal S} (\theta_{m_1}) P_{m_1}^z
  \label{tryuntilproj}
\end{align}
where the product symbol multiplies the matrices in reverse order, i.e. from $ l =1 $ to $ l = L $ from right to left. The unitary rotation is applied on the set $ {\cal S} $ involving half of the spins.  The record of all the measurement outcomes is specified by 
\begin{align}
\vec{m} = (m_1, m_2, \dots, m_L)
\end{align}
In the sequence (\ref{tryuntilproj}), $ L $ is taken large enough that the $ m = 0 $ outcome is obtained on the $L$th projection. For all $ l \in [1,L-1] $, $ m_l \ne 0 $.   For this reason, the final projection always takes the value $ m_L = 0 $, and the final unitary rotation is simply the identity matrix.  

The angles $ \theta_m $  are chosen such as to maximize the probability of getting a $ m = 0 $ outcome on the subsequent projection. That is, the state after a projection to spin $ m $ is rotated to
\begin{align}
  U^y_{\cal S}(\theta_m) |J,d, m \rangle =  &  \sum_{J' d' m'}  |J',d', m' \rangle 
   \nonumber \\
  &  \times \langle J', d', m'|  U^y_{\cal S}(\theta_m) | J,d, m \rangle.  
    \label{rotationdicke}
\end{align}
The rotation operation repopulates other Dicke states by putting them in linear superposition. A suitable choice for the rotation angle is obtained by 
making the magnitude of the matrix element for $ m' =0 $ large, which we take to be
\begin{equation}
\label{eq:thetadependence}
\theta_m =  \arcsin \left(\frac{ m }{\sqrt{J_\mathrm{max}(J_\mathrm{max}+1)}} \right) .
\end{equation}
For $m =0$,  $\theta_m = 0$  and the rotation operation is the identity matrix.  It is important that the rotation is performed only on half the spins as otherwise the rotation (\ref{rotationdicke}) is not able to change $ J $, i.e. only $ J' = J $ are non-zero in the matrix elements.  Since our aim is to obtain the singlet state $ J = 0 $, the ability to change $ J $ is obviously crucial to the operation.  Which spins are chosen do not particularly matter for the procedure, they may be chosen at random.  

Once the desired $ m = 0 $ outcome is obtained, another repeat-until-success projection sequence is applied, this time in the $ x $-basis
\begin{align}
{\cal P}^x_{\vec{m}} & = \prod_{l=1}^L [U^z_{\cal S} (\theta_{m_l}) P_{m_l}^x ] \nonumber \\
& =  P_0^x U^z_{\cal S} (\theta_{m_{L-1}}) P_{m_{L-1}}^x \dots U^z_{\cal S} (\theta_{m_1}) P_{m_1}^x .
\label{tryuntilsuccessx}
\end{align}
Again, the unitary operations are designed such as the probability of obtaining $ m = 0 $ on the next measurement is maximized.  

The final sequence that we propose, in the pure states formalism is (see Fig. \ref{figflow}) 
\begin{align}
 \prod_{k=1}^{K-1} ( {\cal P}^x_{\vec{m}_{2k}} {\cal P}^z_{\vec{m}_{2k-1}}  ) | \psi_0 \rangle \rightarrow \sum_d \psi_d | S_{N,d} \rangle  .  
 \label{purestatefinalalgo}
\end{align}
Here $ \vec{m}_{l} $ is the measurement sequence for the $ l$th round of measurements.  The two measurement sequences are repeated many times until  measurement convergence is attained. Once the singlet state is obtained,  no unitary rotation is necessary and projectors in both the $ z $ and $ x $ bases give $ m = 0 $, as in (\ref{singletsequence}).  We define convergence being attained when the projection sequence such as in (\ref{singletsequence}) consecutively returns $ m = 0 $ for several measurements.   Our claim is that such a sequence always converges to a singlet state, from an arbitrary initial state $ | \psi_0 \rangle $.  On the right hand side of (\ref{purestatefinalalgo}) we have written as an arbitrary superposition with coefficients $ \psi_d  $ of the distinct supersinglet states.  As we shall see, our procedure is sensitive only to whether the state is an a supersinglet state or not, and does not distinguish between the distinct supersinglet states.  

We may also write (\ref{purestatefinalalgo}) for the mixed state case
\begin{align}
& \Big[ \prod_{k=1}^{K-1} ( {\cal P}^x_{\vec{m}_{2k}} {\cal P}^z_{\vec{m}_{2k-1}}  )  \Big] \rho_0  
\Big[ \prod_{k=1}^{K-1} ( {\cal P}^x_{\vec{m}_{2k}} {\cal P}^z_{\vec{m}_{2k-1}}  )  \Big]^\dagger   \nonumber \\
& = {\cal P}^x_{\vec{m}_{2K}} {\cal P}^z_{\vec{m}_{2K-1}} \dots 
  {\cal P}^x_{\vec{m}_{2}} {\cal P}^z_{\vec{m}_{1}}
  \rho_0   {\cal P}^z_{\vec{m}_{1}}   {\cal P}^x_{\vec{m}_{2}} \dots {\cal P}^z_{\vec{m}_{2K-1}}  {\cal P}^x_{\vec{m}_{2K}} 
  \nonumber \\
& \rightarrow \sum_{d d'} \rho_{d d'} |S_{N,d} \rangle \langle S_{N,d'} | .
\label{mixedstateprocedure}
\end{align}
Again, our procedure deterministically generates an arbitrary mixture or superposition of supersinglet states, hence we have written these coefficients as $ \rho_{d d'} $.   

We note that our procedure can be viewed as an adaptation of the imaginary time evolution procedure as presented in Ref. \cite{mao2022measurement}.  In Ref. \cite{mao2022measurement}, a procedure was introduced to target the ground state of a given Hamiltonian.  The procedure involves a sequence of measurements and conditional unitary rotations, chosen in such a way that the target state is a fixed point of the evolution.  This same structure is apparent in the repeat-until-success projection sequence (\ref{tryuntilproj}) and (\ref{tryuntilsuccessx}).  The singlet state is a fixed point of the total sequence as given in (\ref{purestatefinalalgo}), where eventually the sequence converges to only $ m = 0 $ outcomes as in (\ref{singletsequence}).


\section{Numerical evolution}\label{sec:Examples}

We now illustrate the singlet state preparation procedure given in Sec. \ref{sec:Algorithm} by performing a numerical evolution of various cases.

\subsection{Simulation details}
\label{sec:simdetails}

The initial state that we will use in most of our numerical evolutions is a completely mixed state
\begin{equation}
		\label{eq:MixedState}
		\rho_{0} = \frac{\mathbbm{1}^{\otimes N}}{2^N}.
\end{equation}
This state can be viewed equivalently as a thermal state at infinite temperature.  This same initial state was used in works such as Ref. \cite{behbood2014} to experimentally target a singlet state.  Although  the supersinglet state can always be generated from an arbitrary initial state in our procedure, we shall use the completely mixed state as our initial state since it is a state that possesses no entanglement or coherence \cite{radhakrishnan2019basis,ma2019operational}.  Since a supersinglet state possesses both entanglement and coherence, convergence towards the supersinglet state shows that our procedure is responsible for creating these quantum properties.  

The evolution sequence is performed by taking the initial state (\ref{eq:MixedState}) in the procedure (\ref{mixedstateprocedure}).  Each projection operation is chosen randomly according to Born probabilities.  We perform this with an accept/reject procedure as given in Appendix \ref{app:metropolis}.  Due to the randomness of measurements, each run of the procedure gives a different evolution. 
The atoms to which the unitary rotations (\ref{subensemblerotation}) are performed are chosen randomly each time the unitary operation is applied, selecting half the spins at random.  
The measurement sequence is performed multiple times, for which we check for the convergence to the singlet state.  We consider convergence to be attained if 5 consecutive measurements in alternating $ z $ and $ x $ bases give the $ m = 0 $ outcome.  

To characterize the state obtained after a measurement in the sequence, we will evaluate several quantities.  The first is the normalized average value of the total spin squared as given in (\ref{casimir}).  
\begin{align}
{\cal \bar{J}}^2 & \equiv \frac{\langle  J^2  \rangle}{J_{\max} (J_{\max} +1)}  \label{jbardef}  \\
& =0 . \hspace{2cm} \text{(singlet)} \nonumber 
\end{align}
The above quantity has an expectation value of zero for a singlet state since the eigenvalue of the Casimir invariant is $ J (J+1) $.  This quantity lies in the interval $ 0 \le {\cal \bar{J}}^2 \le 1 $ due to the normalization factor.  Only the singlet states have a zero expectation value for this operator and hence it is a good detector for the supersinglet state. The expectation values of $ J^2 $ is non-zero for the thermal state since it involves contributions from total spin sectors $ J > 0 $.  For the first order spin expectation values, both the initial thermal state (\ref{eq:MixedState}) and the singlet state have zero expectation values
\begin{align}
\langle J^\alpha  \rangle =0, \hspace{1cm} \text{(singlet \& completely mixed)}
\end{align}
for $ \alpha \in \{ x, y, z \} $.  Hence the variance of the spins for a singlet state will be zero  
\begin{align}
(\Delta J^\alpha)^2 & \equiv \langle (J^\alpha)^2 \rangle - \langle J^\alpha \rangle^2 \label{spinvars} \\
& = 0 . \hspace{2cm} \text{(singlet)}  \nonumber 
\end{align}
This also shows the squeezed nature of the spin observables for the singlet state.  The other way we will quantify the state is using fidelity, defined as 
\begin{align}
	\label{eq:Fidelity}
F & =\sum_d F_d \\
F_d & = \langle S_{N,d} | \rho | S_{N,d}  \rangle 
\label{eq:Fidelity2}
\end{align}
where $ \rho $ is the state at a particular point in the projection sequence.
This is a useful measure to show the exact state that has been reached in the procedure, whereas (\ref{jbardef}) does not distinguish between distinct supersinglets.  $ F = 1 $ indicates that the state $\rho$ is a singlet state, with $ F_d $ showing the decompositions.

	\subsection{$ N = 4 $ case}\label{sec:sec:FourParticle}
	
We first consider a relatively small system consisting of $ N = 4 $ spins, each with $ j = 1/2 $.  This example will illustrate some of the basic properties of the supersinglet state preparation procedure.   For this and the next section we shall only consider the even $ N $ case, and discuss the odd $ N $ case in Sec. \ref{sec:oddncase}.  

The $N = 4$ case is the smallest system that illustrates that the supersinglet state is not a unique state, due to angular momentum addition.  There are two distinct supersinglet states  \cite{schimpf2010} for four $ j = 1/2 $ spins, given by \cite{cabello2003supersinglets}
\begin{align}
| S_{4,1} \rangle & =  \frac{1}{\sqrt{3}}\left[ \lvert \phi_+\rangle^{\otimes 2} - \lvert\phi_-\rangle^{\otimes 2}  - \lvert\psi_+\rangle^{\otimes 2} \right]  \label{eq:FourParticleSinglet} \\
| S_{4,2} \rangle & = \lvert\psi_-\rangle^{\otimes 2} ,
\label{eq:FourParticleSinglet2}
\end{align}
where we defined the Bell states for two $ j = 1/2 $ spins as 
\begin{equation}
		\label{eq:BellBasis}
		\begin{split}
			\lvert \phi_{\pm}  \rangle & = \frac{1}{\sqrt{2}} \left(\lvert +\rangle | +\rangle \pm  \lvert- \rangle | -\rangle\right) \\
			\lvert \psi_{\pm} \rangle & = \frac{1}{\sqrt{2}} \left(\lvert + \rangle | -\rangle \pm  \lvert- \rangle | +\rangle\right)  .  
		\end{split}
\end{equation} 
Here the $ N = 2 $ singlet state is $ 	\lvert \psi_{-} \rangle  $  and the remaining three states form a $ J = 1$ triplet.  The first $ N = 4 $  supersinglet (\ref{eq:FourParticleSinglet}) is given as linear combination of the triplet states, while the second supersinglet (\ref{eq:FourParticleSinglet2}) is simply two $ N = 2 $ singlet states. Any linear combination or mixture of the two distinct supersinglet states is also a supersinglet state.  The supersinglet state is invariant under rotations of the total spin, as given in (\ref{rotationinvar}).  

	\begin{figure}[t]
		\includegraphics[width=\columnwidth]{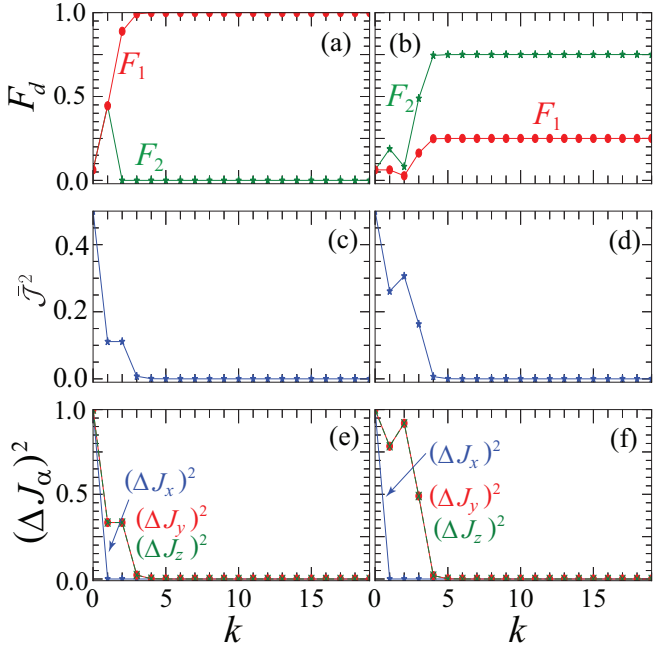}
		\caption{Evolution according to the procedure (\ref{mixedstateprocedure}) for $ N = 4 $ spin $ j = 1/2 $ particles for an initial state being a completely mixed state (\ref{eq:MixedState}). Horizontal axis shows the number of rounds of $ z $ and $ x $-basis measurement sequences, i.e. the variable $ k $ in (\ref{purestatefinalalgo}).   Two different trajectories are shown, for each run the (a)(b) fidelity (\ref{eq:Fidelity2}); (c)(d) normalized total spin squared (\ref{jbardef}); and (e)(f) spin variances (\ref{spinvars}) are shown.  Lines connecting the markers are to guide the eye.}
		\label{fig4}
	\end{figure}
	
Our numerical results are presented in Fig.~\ref{fig4} for two sample evolutions.  We observe from the fidelity and the total spin squared that the supersinglet state is reached after $ \sim 4 $ rounds of projections. The two evolutions that we show are merely examples, and We have run our procedure for over $ 1000 $ trajectories and found that in $100$ \% of the cases the supersinglet state is reached.  The number of projections within each repeat-until-success projection sequence  decreases with the number of rounds, with the largest number being in the first round $k = 1 $.  A typical number of  measurements in the first round is order of $ 10 $ \emph{z}-measurements or \emph{x}-measurements.   While a supersinglet state is always reached with $ F = 1$, Fig.~\ref{fig4}(a)(b) reveals that the final contributions to the two distinct supersinglets are not always the same.  We observe that each time the fidelity $F_1> F_2$, eventually the state approaches the state $ | S_{4,1} \rangle $.  On the hand, whenever $F_2 > F_1$, the fidelities converge towards $ F_1 = 0.25 $ and $ F_2 = 0.75$.  

All spin variances start out being unity corresponding to their value for a completely mixed state. After one round of measurements, the final projection that is made is $ P^x_0 $.  Hence the state of the system after one round of measurements is always an eigenstate of the $ J^x $ operator.  For this reason,  the error of the state for all $k > 0 $ has zero variance for  $J^x $ as seen in Figs.~\ref{fig4}(e)(f).  The variance for the other spin directions are still typically non-zero at this point at $ k = 1$.  For each sequence, the measurement outcomes are random, hence the evolution of the variance evolves towards zero along different trajectories. Another source of randomness is the random selection of the spins to be rotated in the conditional unitary rotations.  As the state settles into a supersinglet state, the error along \emph{y}-axis and \emph{z}-axis becomes zero, as seen in Figs.~\ref{fig4}(e)(f), heralding the realization of a supersinglet state.

\subsection{Spin polarized initial state}

	\begin{figure}[t]
		\includegraphics[width=\columnwidth]{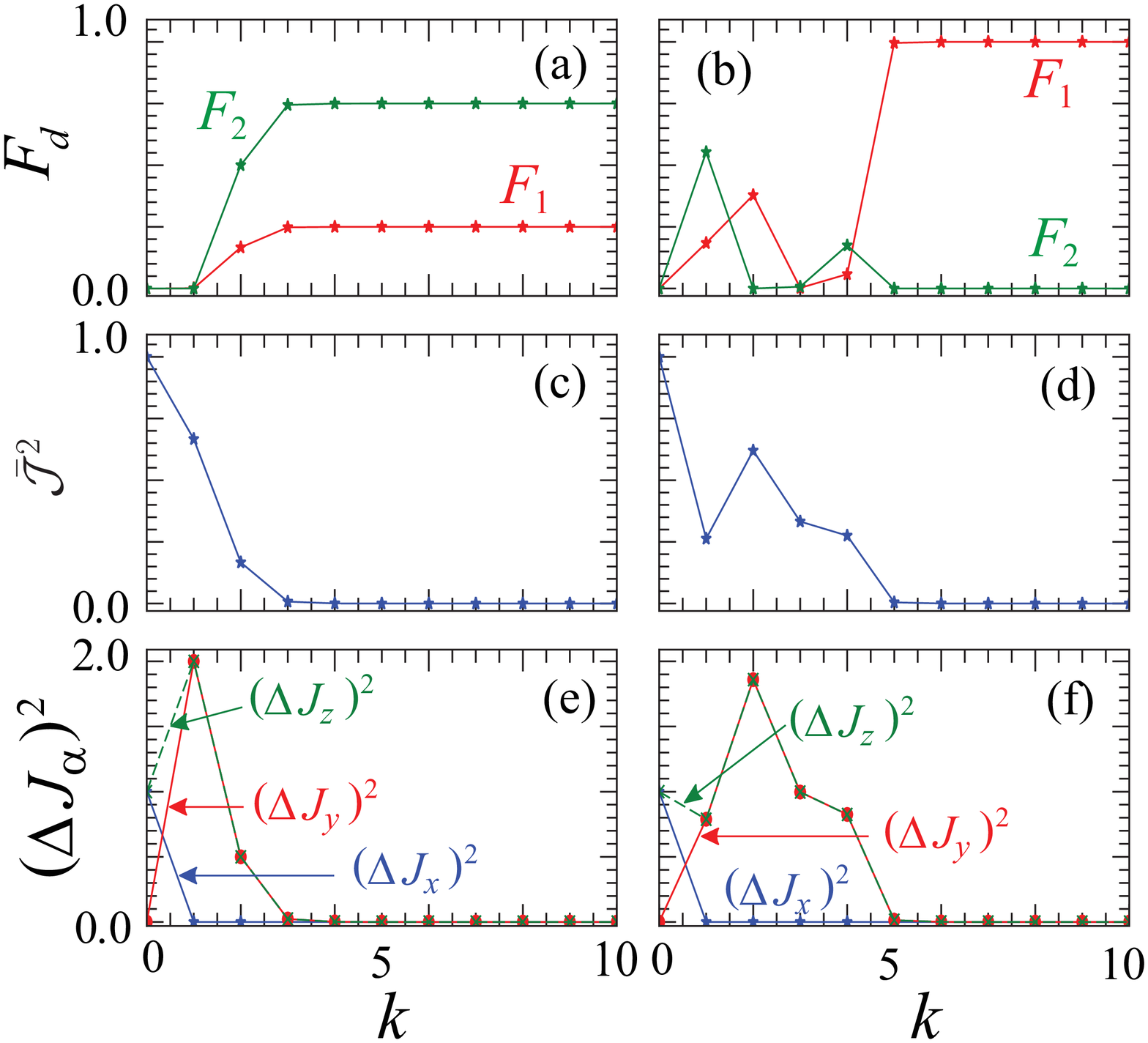}
		\caption{Evolution according to the procedure (\ref{mixedstateprocedure}) for $ N = 4$ spin $ j = 1/2 $ particles with an initial state being a \emph{y}-polarized spin coherent state (\ref{eq:coherentstate}). Horizontal axis shows the number of rounds of $ z $ and $ x $-basis measurement sequences, i.e. the variable $ k $ in (\ref{purestatefinalalgo}).  Two different runs are shown, for each run the (a)(b) fidelity (\ref{eq:Fidelity2}); (c)(d) normalized total spin squared (\ref{jbardef}); and (e)(f) spin variances (\ref{spinvars}) are shown.  Lines connecting the markers are to guide the eye.}
		\label{fig3}
	\end{figure}


Our procedure to generate the supersinglet state is invariant to  initial conditions.  To show this, we now repeat the calculations for $N=4$ and $ j = 1/2$ again, but starting in a spin coherent state \cite{arecchi1972}.  We consider in particular the state polarized in the $ y $-direction
\begin{equation}
	\label{eq:coherentstate}
\lvert \psi_0 \rangle = e^{-i J^x \pi/2}\lvert J,d=1,m=J\rangle .  
\end{equation}
Here $\lvert J,d=1,m=J \rangle$ is the Dicke state polarized in the $ z$-direction. Rotating this around the $ x $-axis by an angle $ \pi/2$ gives a $ y $-polarized state. We choose a $ y $-polarized state since we perform projections in the $ x $ and $ z $ basis, and this produces a large backaction in either case.


Our numerical results are shown in Fig.~\ref{fig3} for two example evolutions. We see a similar behavior to the results obtained in Fig.~\ref{fig4}.  We observe from the fidelity and the total spin squared that the supersinglet state is reached after $ \sim 5 $ rounds of projections. While a supersinglet state is always reached with $ F = 1$, Fig.~\ref{fig3}(a)(b) reveals that the final contributions to the two distinct supersinglets are not always the same. Again the same pattern of convergence to either $ F_1 = 0.25, F_2 = 0.75$ or $ F_1 = 1, F_2 = 0$ is seen as with the previous case. For the variances, since the initial state is polarized along \emph{y}-axis, the $J^y$ spin variance starts out being zero, and the variances of $J^x, J^z$ start out being unity.  After one round of measurements, the final projection that is made is $ P^x_0 $.  For this reason, the error of the state for all $k > 0 $ has zero variance for $J^x $ as seen in Figs.~\ref{fig3}(e)(f).  The variance of $J^y$ increases after the first measurement to the same value as the variance of $J^z$, due to the effect of backaction.   As the state settles into a supersinglet state, the error along \emph{y}-axis and \emph{z}-axis becomes zero, as seen in Figs.~\ref{fig4}(e)(f).

\subsection{Larger number of spins}

We now repeat the calculation for a larger number of spins. We consider $ N $ spin-$1/2$ particles, where we take $ N $ even.  Due to the necessity to simulate the full Hilbert space of dimension $ 2^N $ the largest system that we could simulate within a reasonable time was $ N = 10 $.   The major numerical overhead results evaluating matrix multiplications due to the unitary transformations, which have a dimension $ 2^N \times 2^N $.  For this case, we will only calculate the expectation values and variances of the spin operators, since there are a larger number of distinct singlet states.  In this case there are a total of $42$ distinct singlet states~\cite{schimpf2010}.  We follow the procedure as given in Sec. \ref{sec:simdetails}, again starting from a completely mixed state.  

\begin{figure}[t]
		\includegraphics[width=\columnwidth]{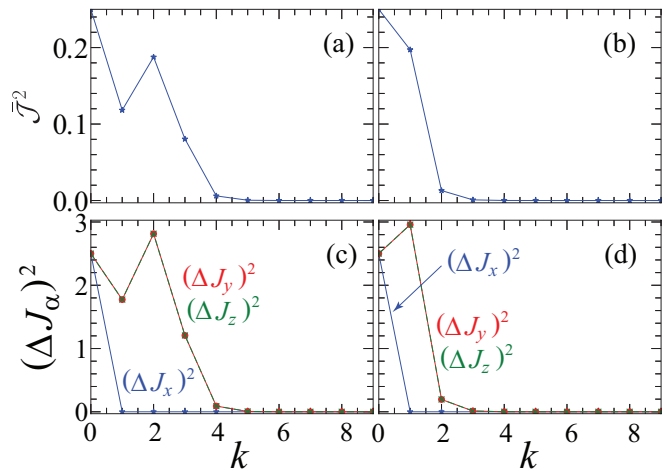}
		\caption{Evolution according to the procedure (\ref{mixedstateprocedure}) for $ N = 10 $ spin $ j = 1/2 $ particles. Horizontal axis shows the number of rounds of $ z $ and $ x $-basis measurement sequences, i.e. the variable $ k $ in (\ref{purestatefinalalgo}).  The initial state state is set as the completely mixed state (\ref{eq:MixedState}). Two different runs are shown, for each run the (a)(b) normalized total spin squared (\ref{jbardef}); and (c)(d) spin variances (\ref{spinvars}) are shown.  Lines connecting the markers are to guide the eye.}
		\label{fig5}
	\end{figure}

The results are shown in Fig. \ref{fig5}.  As for the smaller system size considered before, the state converges towards a supersinglet state, as can be seen from the total spin squared operator.  Remarkably, the convergence is attained with a similar number of rounds of measurement sequences, typically after about $ k = 5$ full convergence was attained.  The number of projections within a single projection sequence was larger, the number of measurements in a single sequence (\ref{tryuntilproj}) and (\ref{tryuntilsuccessx}) was typically of the order of $ 100 $  \emph{z}-measurements and \emph{x}-measurements.   It is to be expected that a larger number of projections are necessary for a larger system size, due to the larger Hilbert space that the state must traverse during the evolution.  It is nevertheless remarkable that such fast convergence is attained for a significantly larger system.  

For the variances, we again see fast convergence towards zero variance for all spin directions.  As was the case for the smaller system, some random fluctuations are seen during the evolution, where occasionally the variance increases.  However, within $ \sim 5 $ rounds variances decay to zero.

\subsection{Odd number of spins}
\label{sec:oddncase}

Depending upon the experimental realization, the number of spins $ N $ may not be a precisely controllable quantity.  This is true of the atomic ensemble implementation that is suggested in Sec. \ref{sec:atomic}, where the number of atoms is typically very large and not controlled at the single atom level.  A potential problem arises here because for odd $ N $ a supersinglet state does not exist.  For example, the smallest total spin that can be realized for an odd number of $ j = 1/2 $ spins is a total spin of $ J = 1/2$.   We show in this section how such a scenario can be handled, and how convergence can still be attained towards a small total spin.

To handle this case, we follow the same procedure as given in Fig. \ref{figflow}, except that we replace the criterion for exiting the repeat-until-success projection sequence to  $\lvert m \rvert \le m_{\text{cut}} $.  Hence for targeting a genuine singlet state we have $ m_{\text{cut}} = 0 $, but for the odd $ N $ case we set $ m_{\text{cut}} = 1/2 $, since $ m = 0 $ does not exist.  This cutoff introduces a wider variety of states that the sequence can potentially converge to. This makes the convergence in the repeat-until-success projection sequence typically faster, since there are more states that are allowable in the criterion.  However, there is a trade-off as a larger $ m_{\text{cut}} $ reduces the fidelity with respect to the desired target state.  $ m_{\text{cut}} $  can be viewed as an adjustable parameter that can control the convergence speed, at the expense of a less accurate target state.  Having such a parameter can be useful even in the context of even $ N $, if $ N $ is a very large number.  A very large $ N $ may mean that the convergence of the repeat-until-success sequence is rather slow, but this can be mitigated by reducing the accuracy by increasing $ m_{\text{cut}} $.  

\begin{figure}[t]
	\includegraphics[width=\columnwidth]{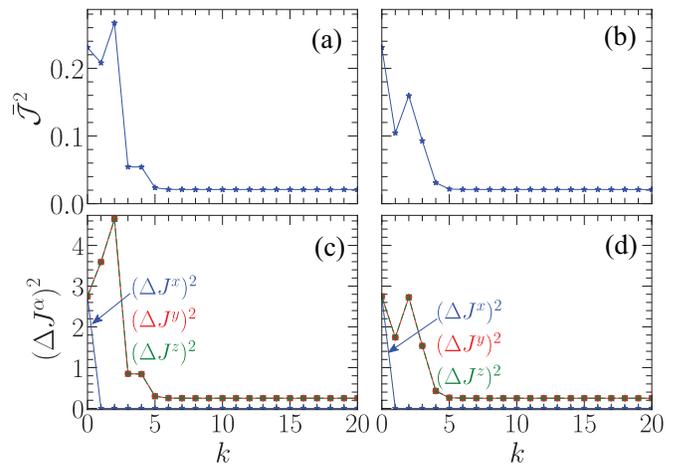}
	\caption{Evolution according to the procedure (\ref{mixedstateprocedure}) for $ N = 11 $ spin $ j = 1/2 $ particles. Horizontal axis shows the number of rounds of $ z $ and $ x $-basis measurement sequences, i.e. the variable $ k $ in (\ref{purestatefinalalgo}).  The initial state state is set as the completely mixed state (\ref{eq:MixedState}). Two different runs are shown, for each run the (a)(b) normalized total spin squared (\ref{jbardef}); and (c)(d) spin variances (\ref{spinvars}) are shown.  Lines connecting the markers are to guide the eye.}
		\label{fig6}
\end{figure}

Our numerical results are shown in Fig. \ref{fig6}.  As with the even $ N$  case considered before, the state converges towards a minimum $J$ state, as can be seen from the total spin squared operator. The minimum state occurs for $J=  1/2$ giving the minimum expectation value of the Casimir invariant as $\langle J^2 \rangle  = 3/4$, which in terms of the normalized values is   $\bar{\mathcal{J}}^2 = 3/143 \approx 0.02$.  The variance of $J_x$ is always zero for the same reasons given previously.  However, the variances of $J_y$ and $J_z$ settle to a minimum value of $0.25$, in a departure from the  even  $N $ case.  Interestingly, the convergence to the minimal total spin squared state is attained with a similar number of rounds of measurement sequences as to the even  \emph{N} case, typically after about $ k = 5$ rounds.  


 \section{Summary and Conclusions \label{sec:Discussion}}

We have proposed a scheme based on a sequence of projective measurements in the collective spin basis and conditional unitary rotations to prepare a supersinglet state. The main procedure is given in (\ref{purestatefinalalgo}) and summarized in Fig. \ref{figflow}.  The scheme deterministically produces the supersinglet state from an arbitrary initial state without postselection.   Within a measurement sequence, the quantum state of the atoms collapses randomly to a state given by the projection operator. Convergence of the state towards the supersinglet is ensured by unitary rotations that repopulate the spin zero population.  Using the property that a supersinglet state is invariant under total spin rotations, repeated projections in the $ z $ and $ x $ bases results in convergence to the spin zero state.  The procedure is compatible for an arbitrary number of particles $ N $ and spin  $ j $.

We have found that the procedure is remarkably efficient in converging to the supersinglet state, with little difference seen in terms of the rounds of measurements required for various $ N $.  Within each measurement sequence (\ref{tryuntilproj}) and (\ref{tryuntilsuccessx}) we note that it does take longer to find the $ m = 0 $ outcome, which is to be expected due to the larger Hilbert space. In an atomic ensemble implementation where the atom numbers can be far larger than that simulated here (e.g. $ N = 10^6 $) one may worry that this will make the convergence excessively long.  The number of iterations can however be alleviated by introducing tolerances to the target state.  As seen in Sec. \ref{sec:oddncase} it is also possible to adjust the procedure such that convergence is towards a range of spins, and not only $ m = 0 $ exactly.   By setting the tolerance to the target spin sector to a larger value, one may reduce the convergence time, at the expense of a lower fidelity of the spin zero sector. In this way, the procedure should be applicable also to macroscopic systems.  

We note that in contrast to existing schemes, our procedure  produces the exact supersinglet state, as opposed to producing squeezing as in Ref. \cite{toth2010generation}.  This can be seen in our results in Figs. \ref{fig4}-\ref{fig5}, where the spin variances reduce to zero for all directions.  The main reason for this is that our measurements are projective, which induce a much more dramatic effect on the state of the atoms, due to the strong measurements that affect the state. This results in a much faster convergence to the supersinglet state.  Another attractive feature of our procedure is that it allows for a deterministic way of preparing the supersinglet state without involving postselection of the final state. There is also no need to prepare a special initial state.  The experimental requirements only involve QND measurements which are routinely performed, and rotations on subensembles of the spins, which should be within current experimental capabilities.

\acknowledgments
This work is supported by the National Natural Science Foundation of China (62071301); NYU-ECNU Institute of Physics at NYU Shanghai; the Joint Physics Research Institute Challenge Grant; the Science and Technology Commission of Shanghai Municipality (19XD1423000,22ZR1444600); the NYU Shanghai Boost Fund; the China Foreign Experts Program (G2021013002L); the NYU Shanghai Major-Grants Seed Fund; and 
the Talented Young Scientists Program (NGA-16-001) supported by the Ministry of Science and Technology of China.
\appendix

\section{Choosing measurement by Born probabilities}
\label{app:metropolis}

Each projective measurement performed in (\ref{tryuntilproj}) and (\ref{tryuntilsuccessx}) must be chosen randomly according to Born probability outcomes.  Here we show how we choose the measurement outcomes using an accept/reject procedure.  

Suppose we wish to perform a projective measurement $     P_m | \phi \rangle  $
where $  P_m $ are projectors and $ | \phi \rangle $ is an arbitrary initial state.  The probability of the $ m$th outcome is given by 
\begin{align}
    p_m = \langle \phi | P_m | \phi \rangle 
    \label{bornprobs}
\end{align}
using the idempotency of projection operators.  Then we may select measurement outcomes according to Born probabilities using the following algorithm. 
\texttt{\begin{itemize}
		\item[1.] Choose a proposed outcome $ m $ randomly from a uniform distribution.  
		\item[2.] Calculate the probability $ p_m $ according to  (\ref{bornprobs}).    
		\item[3.] Choose a random real number $ r $ in the interval  $ [0,1] $ from a uniform distribution. 
		\item[4.] If $ r < p_m $ then reject and go to step 1.  Otherwise accept $ m $ as the measurement outcome.  
\end{itemize} }


\begin{thebibliography}{62}%
	\makeatletter
	\providecommand \@ifxundefined [1]{%
		\@ifx{#1\undefined}
	}%
	\providecommand \@ifnum [1]{%
		\ifnum #1\expandafter \@firstoftwo
		\else \expandafter \@secondoftwo
		\fi
	}%
	\providecommand \@ifx [1]{%
		\ifx #1\expandafter \@firstoftwo
		\else \expandafter \@secondoftwo
		\fi
	}%
	\providecommand \natexlab [1]{#1}%
	\providecommand \enquote  [1]{``#1''}%
	\providecommand \bibnamefont  [1]{#1}%
	\providecommand \bibfnamefont [1]{#1}%
	\providecommand \citenamefont [1]{#1}%
	\providecommand \href@noop [0]{\@secondoftwo}%
	\providecommand \href [0]{\begingroup \@sanitize@url \@href}%
	\providecommand \@href[1]{\@@startlink{#1}\@@href}%
	\providecommand \@@href[1]{\endgroup#1\@@endlink}%
	\providecommand \@sanitize@url [0]{\catcode `\\12\catcode `\$12\catcode
		`\&12\catcode `\#12\catcode `\^12\catcode `\_12\catcode `\%12\relax}%
	\providecommand \@@startlink[1]{}%
	\providecommand \@@endlink[0]{}%
	\providecommand \url  [0]{\begingroup\@sanitize@url \@url }%
	\providecommand \@url [1]{\endgroup\@href {#1}{\urlprefix }}%
	\providecommand \urlprefix  [0]{URL }%
	\providecommand \Eprint [0]{\href }%
	\providecommand \doibase [0]{http://dx.doi.org/}%
	\providecommand \selectlanguage [0]{\@gobble}%
	\providecommand \bibinfo  [0]{\@secondoftwo}%
	\providecommand \bibfield  [0]{\@secondoftwo}%
	\providecommand \translation [1]{[#1]}%
	\providecommand \BibitemOpen [0]{}%
	\providecommand \bibitemStop [0]{}%
	\providecommand \bibitemNoStop [0]{.\EOS\space}%
	\providecommand \EOS [0]{\spacefactor3000\relax}%
	\providecommand \BibitemShut  [1]{\csname bibitem#1\endcsname}%
	\let\auto@bib@innerbib\@empty
	\bibitem [{\citenamefont {Anderson}(1987)}]{anderson1987}%
	\BibitemOpen
	\bibfield  {author} {\bibinfo {author} {\bibfnamefont {P.~W.}\ \bibnamefont
			{Anderson}},\ }\bibfield  {title} {\enquote {\bibinfo {title} {{The
					Resonating Valence Bond State in La\textsubscript{2}CuO\textsubscript{4} and
					Superconductivity}},}\ }\href {\doibase 10.1126/science.235.4793.1196}
	{\bibfield  {journal} {\bibinfo  {journal} {Science}\ }\textbf {\bibinfo
			{volume} {235}},\ \bibinfo {pages} {1196--1198} (\bibinfo {year}
		{1987})}\BibitemShut {NoStop}%
	\bibitem [{\citenamefont {Werlang}\ \emph {et~al.}(2010)\citenamefont
		{Werlang}, \citenamefont {Trippe}, \citenamefont {Ribeiro},\ and\
		\citenamefont {Rigolin}}]{werlang2010}%
	\BibitemOpen
	\bibfield  {author} {\bibinfo {author} {\bibfnamefont {T.}~\bibnamefont
			{Werlang}}, \bibinfo {author} {\bibfnamefont {C.}~\bibnamefont {Trippe}},
		\bibinfo {author} {\bibfnamefont {G.~A.~P.}\ \bibnamefont {Ribeiro}}, \ and\
		\bibinfo {author} {\bibfnamefont {G.}~\bibnamefont {Rigolin}},\ }\bibfield
	{title} {\enquote {\bibinfo {title} {{Quantum Correlations in Spin Chains at
					Finite Temperatures and Quantum Phase Transitions}},}\ }\href {\doibase
		10.1103/PhysRevLett.105.095702} {\bibfield  {journal} {\bibinfo  {journal}
			{Phys. Rev. Lett.}\ }\textbf {\bibinfo {volume} {105}},\ \bibinfo {pages}
		{095702} (\bibinfo {year} {2010})}\BibitemShut {NoStop}%
	\bibitem [{\citenamefont {Sachdev}(2011)}]{sachdev2011}%
	\BibitemOpen
	\bibfield  {author} {\bibinfo {author} {\bibfnamefont {S.}~\bibnamefont
			{Sachdev}},\ }\href@noop {} {\emph {\bibinfo {title} {Quantum Phase
				Transitions}}},\ \bibinfo {edition} {2nd}\ ed.\ (\bibinfo  {publisher}
	{Cambridge University Press},\ \bibinfo {address} {Cambridge},\ \bibinfo
	{year} {2011})\BibitemShut {NoStop}%
	\bibitem [{\citenamefont {Huber}\ \emph {et~al.}(2015)\citenamefont {Huber},
		\citenamefont {Perarnau-Llobet}, \citenamefont {Hovhannisyan}, \citenamefont
		{Skrzypczyk}, \citenamefont {Kl{\"o}ckl}, \citenamefont {Brunner},\ and\
		\citenamefont {Ac{\'i}n}}]{huber2015}%
	\BibitemOpen
	\bibfield  {author} {\bibinfo {author} {\bibfnamefont {M.}~\bibnamefont
			{Huber}}, \bibinfo {author} {\bibfnamefont {M.}~\bibnamefont
			{Perarnau-Llobet}}, \bibinfo {author} {\bibfnamefont {K.~V.}\ \bibnamefont
			{Hovhannisyan}}, \bibinfo {author} {\bibfnamefont {P.}~\bibnamefont
			{Skrzypczyk}}, \bibinfo {author} {\bibfnamefont {C.}~\bibnamefont
			{Kl{\"o}ckl}}, \bibinfo {author} {\bibfnamefont {N.}~\bibnamefont {Brunner}},
		\ and\ \bibinfo {author} {\bibfnamefont {A.}~\bibnamefont {Ac{\'i}n}},\
	}\bibfield  {title} {\enquote {\bibinfo {title} {{Thermodynamic cost of
					creating correlations}},}\ }\href@noop {} {\bibfield  {journal} {\bibinfo
			{journal} {New J. Phys.}\ }\textbf {\bibinfo {volume} {17}},\ \bibinfo
		{pages} {065008} (\bibinfo {year} {2015})}\BibitemShut {NoStop}%
	\bibitem [{\citenamefont {Nielsen}\ and\ \citenamefont
		{Chuang}(2000)}]{nielsen2000}%
	\BibitemOpen
	\bibfield  {author} {\bibinfo {author} {\bibfnamefont {M.~A.}\ \bibnamefont
			{Nielsen}}\ and\ \bibinfo {author} {\bibfnamefont {I.~L.}\ \bibnamefont
			{Chuang}},\ }\href@noop {} {\emph {\bibinfo {title} {Quantum computation and
				quantum information}}}\ (\bibinfo  {publisher} {Cambridge University Press},\
	\bibinfo {address} {Cambridge},\ \bibinfo {year} {2000})\BibitemShut
	{NoStop}%
	\bibitem [{\citenamefont {Bloch}\ \emph {et~al.}(2012)\citenamefont {Bloch},
		\citenamefont {Dalibard},\ and\ \citenamefont {Nascimb{\`e}ne}}]{bloch2012}%
	\BibitemOpen
	\bibfield  {author} {\bibinfo {author} {\bibfnamefont {I.}~\bibnamefont
			{Bloch}}, \bibinfo {author} {\bibfnamefont {J.}~\bibnamefont {Dalibard}}, \
		and\ \bibinfo {author} {\bibfnamefont {S.}~\bibnamefont {Nascimb{\`e}ne}},\
	}\bibfield  {title} {\enquote {\bibinfo {title} {{Quantum simulations with
					ultracold quantum gases}},}\ }\href {\doibase doi.org/10.1038/nphys2259}
	{\bibfield  {journal} {\bibinfo  {journal} {Nature Phys}\ }\textbf {\bibinfo
			{volume} {8}},\ \bibinfo {pages} {267--276} (\bibinfo {year}
		{2012})}\BibitemShut {NoStop}%
	\bibitem [{\citenamefont {Georgescu}\ \emph {et~al.}(2014)\citenamefont
		{Georgescu}, \citenamefont {Ashhab},\ and\ \citenamefont
		{Nori}}]{georgescu2014}%
	\BibitemOpen
	\bibfield  {author} {\bibinfo {author} {\bibfnamefont {I.~M.}\ \bibnamefont
			{Georgescu}}, \bibinfo {author} {\bibfnamefont {S.}~\bibnamefont {Ashhab}}, \
		and\ \bibinfo {author} {\bibfnamefont {F.}~\bibnamefont {Nori}},\ }\bibfield
	{title} {\enquote {\bibinfo {title} {{Quantum simulation}},}\ }\href
	{\doibase 10.1103/RevModPhys.86.153} {\bibfield  {journal} {\bibinfo
			{journal} {Rev. Mod. Phys.}\ }\textbf {\bibinfo {volume} {86}},\ \bibinfo
		{pages} {153--185} (\bibinfo {year} {2014})}\BibitemShut {NoStop}%
	\bibitem [{\citenamefont {Afzelius}\ \emph {et~al.}(2015)\citenamefont
		{Afzelius}, \citenamefont {Gisin},\ and\ \citenamefont
		{de~Riedmatten}}]{afzelius2015}%
	\BibitemOpen
	\bibfield  {author} {\bibinfo {author} {\bibfnamefont {M.}~\bibnamefont
			{Afzelius}}, \bibinfo {author} {\bibfnamefont {N.}~\bibnamefont {Gisin}}, \
		and\ \bibinfo {author} {\bibfnamefont {H.}~\bibnamefont {de~Riedmatten}},\
	}\bibfield  {title} {\enquote {\bibinfo {title} {{Quantum memory for
					photons}},}\ }\href@noop {} {\bibfield  {journal} {\bibinfo  {journal}
			{Physics Today}\ }\textbf {\bibinfo {volume} {68}},\ \bibinfo {pages}
		{42--47} (\bibinfo {year} {2015})}\BibitemShut {NoStop}%
	\bibitem [{\citenamefont {Byrnes}\ \emph {et~al.}(2007)\citenamefont {Byrnes},
		\citenamefont {Recher}, \citenamefont {Kim}, \citenamefont {Utsunomiya},\
		and\ \citenamefont {Yamamoto}}]{byrnes2007quantum}%
	\BibitemOpen
	\bibfield  {author} {\bibinfo {author} {\bibfnamefont {Tim}\ \bibnamefont
			{Byrnes}}, \bibinfo {author} {\bibfnamefont {Patrik}\ \bibnamefont {Recher}},
		\bibinfo {author} {\bibfnamefont {Na~Young}\ \bibnamefont {Kim}}, \bibinfo
		{author} {\bibfnamefont {Shoko}\ \bibnamefont {Utsunomiya}}, \ and\ \bibinfo
		{author} {\bibfnamefont {Yoshihisa}\ \bibnamefont {Yamamoto}},\ }\bibfield
	{title} {\enquote {\bibinfo {title} {Quantum simulator for the hubbard model
				with long-range coulomb interactions using surface acoustic waves},}\
	}\href@noop {} {\bibfield  {journal} {\bibinfo  {journal} {Physical review
				letters}\ }\textbf {\bibinfo {volume} {99}},\ \bibinfo {pages} {016405}
		(\bibinfo {year} {2007})}\BibitemShut {NoStop}%
	\bibitem [{\citenamefont {Giovannetti}\ \emph {et~al.}(2004)\citenamefont
		{Giovannetti}, \citenamefont {Lloyd},\ and\ \citenamefont
		{Maccone}}]{giovannetti2004}%
	\BibitemOpen
	\bibfield  {author} {\bibinfo {author} {\bibfnamefont {V.}~\bibnamefont
			{Giovannetti}}, \bibinfo {author} {\bibfnamefont {S.}~\bibnamefont {Lloyd}},
		\ and\ \bibinfo {author} {\bibfnamefont {L.}~\bibnamefont {Maccone}},\
	}\bibfield  {title} {\enquote {\bibinfo {title} {{Quantum-Enhanced
					Measurements: Beating the Standard Quantum Limit}},}\ }\href@noop {}
	{\bibfield  {journal} {\bibinfo  {journal} {Science}\ }\textbf {\bibinfo
			{volume} {306}},\ \bibinfo {pages} {1330--1336} (\bibinfo {year}
		{2004})}\BibitemShut {NoStop}%
	\bibitem [{\citenamefont {You}\ \emph {et~al.}(2017)\citenamefont {You},
		\citenamefont {Adhikari}, \citenamefont {Chi}, \citenamefont {LaBorde},
		\citenamefont {Matyas}, \citenamefont {Zhang}, \citenamefont {Su},
		\citenamefont {Byrnes}, \citenamefont {Lu}, \citenamefont {Dowling} \emph
		{et~al.}}]{you2017multiparameter}%
	\BibitemOpen
	\bibfield  {author} {\bibinfo {author} {\bibfnamefont {Chenglong}\
			\bibnamefont {You}}, \bibinfo {author} {\bibfnamefont {Sushovit}\
			\bibnamefont {Adhikari}}, \bibinfo {author} {\bibfnamefont {Yuxi}\
			\bibnamefont {Chi}}, \bibinfo {author} {\bibfnamefont {Margarite~L}\
			\bibnamefont {LaBorde}}, \bibinfo {author} {\bibfnamefont {Corey~T}\
			\bibnamefont {Matyas}}, \bibinfo {author} {\bibfnamefont {Chenyu}\
			\bibnamefont {Zhang}}, \bibinfo {author} {\bibfnamefont {Zuen}\ \bibnamefont
			{Su}}, \bibinfo {author} {\bibfnamefont {Tim}\ \bibnamefont {Byrnes}},
		\bibinfo {author} {\bibfnamefont {Chaoyang}\ \bibnamefont {Lu}}, \bibinfo
		{author} {\bibfnamefont {Jonathan~P}\ \bibnamefont {Dowling}},  \emph
		{et~al.},\ }\bibfield  {title} {\enquote {\bibinfo {title} {Multiparameter
				estimation with single photons—linearly-optically generated quantum
				entanglement beats the shotnoise limit},}\ }\href@noop {} {\bibfield
		{journal} {\bibinfo  {journal} {Journal of Optics}\ }\textbf {\bibinfo
			{volume} {19}},\ \bibinfo {pages} {124002} (\bibinfo {year}
		{2017})}\BibitemShut {NoStop}%
	\bibitem [{\citenamefont {Urizar-Lanz}\ \emph {et~al.}(2013)\citenamefont
		{Urizar-Lanz}, \citenamefont {Hyllus}, \citenamefont {Egusquiza},
		\citenamefont {Mitchell},\ and\ \citenamefont {T{\'o}th}}]{urizar-lanz2013}%
	\BibitemOpen
	\bibfield  {author} {\bibinfo {author} {\bibfnamefont {I.}~\bibnamefont
			{Urizar-Lanz}}, \bibinfo {author} {\bibfnamefont {P.}~\bibnamefont {Hyllus}},
		\bibinfo {author} {\bibfnamefont {I.~L.}\ \bibnamefont {Egusquiza}}, \bibinfo
		{author} {\bibfnamefont {M.~W.}\ \bibnamefont {Mitchell}}, \ and\ \bibinfo
		{author} {\bibfnamefont {G.}~\bibnamefont {T{\'o}th}},\ }\bibfield  {title}
	{\enquote {\bibinfo {title} {{Macroscopic singlet states for gradient
					magnetometry}},}\ }\href {\doibase 10.1103/PhysRevA.88.013626} {\bibfield
		{journal} {\bibinfo  {journal} {Phys. Rev. A}\ }\textbf {\bibinfo {volume}
			{88}},\ \bibinfo {pages} {013626} (\bibinfo {year} {2013})}\BibitemShut
	{NoStop}%
	\bibitem [{\citenamefont {Jing}\ \emph {et~al.}(2019)\citenamefont {Jing},
		\citenamefont {Fadel}, \citenamefont {Ivannikov},\ and\ \citenamefont
		{Byrnes}}]{jing2019}%
	\BibitemOpen
	\bibfield  {author} {\bibinfo {author} {\bibfnamefont {Y.}~\bibnamefont
			{Jing}}, \bibinfo {author} {\bibfnamefont {M.}~\bibnamefont {Fadel}},
		\bibinfo {author} {\bibfnamefont {V.}~\bibnamefont {Ivannikov}}, \ and\
		\bibinfo {author} {\bibfnamefont {T.}~\bibnamefont {Byrnes}},\ }\bibfield
	{title} {\enquote {\bibinfo {title} {Split spin-squeezed {B}ose-{E}instein
				condensates},}\ }\href@noop {} {\bibfield  {journal} {\bibinfo  {journal}
			{New J. Phys.}\ }\textbf {\bibinfo {volume} {21}},\ \bibinfo {pages} {093038}
		(\bibinfo {year} {2019})}\BibitemShut {NoStop}%
	\bibitem [{\citenamefont {Jozsa}\ \emph {et~al.}(2000)\citenamefont {Jozsa},
		\citenamefont {Abrams}, \citenamefont {Dowling},\ and\ \citenamefont
		{Williams}}]{jozsa2000}%
	\BibitemOpen
	\bibfield  {author} {\bibinfo {author} {\bibfnamefont {R.}~\bibnamefont
			{Jozsa}}, \bibinfo {author} {\bibfnamefont {D.~S.}\ \bibnamefont {Abrams}},
		\bibinfo {author} {\bibfnamefont {J.~P.}\ \bibnamefont {Dowling}}, \ and\
		\bibinfo {author} {\bibfnamefont {C.~P.}\ \bibnamefont {Williams}},\
	}\bibfield  {title} {\enquote {\bibinfo {title} {Quantum clock
				synchronisation based on shared prior entanglement},}\ }\href@noop {}
	{\bibfield  {journal} {\bibinfo  {journal} {Phys. Rev. Lett.}\ }\textbf
		{\bibinfo {volume} {85}},\ \bibinfo {pages} {2010--2013} (\bibinfo {year}
		{2000})}\BibitemShut {NoStop}%
	\bibitem [{\citenamefont {Ilo-Okeke}\ \emph {et~al.}(2018)\citenamefont
		{Ilo-Okeke}, \citenamefont {Tessler}, \citenamefont {Dowling},\ and\
		\citenamefont {Byrnes}}]{ilo-okeke2018}%
	\BibitemOpen
	\bibfield  {author} {\bibinfo {author} {\bibfnamefont {E.~O.}\ \bibnamefont
			{Ilo-Okeke}}, \bibinfo {author} {\bibfnamefont {L.}~\bibnamefont {Tessler}},
		\bibinfo {author} {\bibfnamefont {J.~P.}\ \bibnamefont {Dowling}}, \ and\
		\bibinfo {author} {\bibfnamefont {T.}~\bibnamefont {Byrnes}},\ }\bibfield
	{title} {\enquote {\bibinfo {title} {Remote quantum clock synchronization
				without synchronized clocks},}\ }\href@noop {} {\bibfield  {journal}
		{\bibinfo  {journal} {npj Quantum Inf}\ }\textbf {\bibinfo {volume} {4}},\
		\bibinfo {pages} {40} (\bibinfo {year} {2018})}\BibitemShut {NoStop}%
	\bibitem [{\citenamefont {Perseguers}\ \emph {et~al.}(2008)\citenamefont
		{Perseguers}, \citenamefont {Cirac}, \citenamefont {Ac\'{\i}n}, \citenamefont
		{Lewenstein},\ and\ \citenamefont {Wehr}}]{perseguers2008}%
	\BibitemOpen
	\bibfield  {author} {\bibinfo {author} {\bibfnamefont {S.}~\bibnamefont
			{Perseguers}}, \bibinfo {author} {\bibfnamefont {J.~I.}\ \bibnamefont
			{Cirac}}, \bibinfo {author} {\bibfnamefont {A.}~\bibnamefont {Ac\'{\i}n}},
		\bibinfo {author} {\bibfnamefont {M.}~\bibnamefont {Lewenstein}}, \ and\
		\bibinfo {author} {\bibfnamefont {J.}~\bibnamefont {Wehr}},\ }\bibfield
	{title} {\enquote {\bibinfo {title} {{Entanglement distribution in pure-state
					quantum networks}},}\ }\href {\doibase 10.1103/PhysRevA.77.022308} {\bibfield
		{journal} {\bibinfo  {journal} {Phys. Rev. A}\ }\textbf {\bibinfo {volume}
			{77}},\ \bibinfo {pages} {022308} (\bibinfo {year} {2008})}\BibitemShut
	{NoStop}%
	\bibitem [{\citenamefont {Broadfoot}\ \emph {et~al.}(2010)\citenamefont
		{Broadfoot}, \citenamefont {Dorner},\ and\ \citenamefont
		{Jaksch}}]{broadfoot2010}%
	\BibitemOpen
	\bibfield  {author} {\bibinfo {author} {\bibfnamefont {S.}~\bibnamefont
			{Broadfoot}}, \bibinfo {author} {\bibfnamefont {U.}~\bibnamefont {Dorner}}, \
		and\ \bibinfo {author} {\bibfnamefont {D.}~\bibnamefont {Jaksch}},\
	}\bibfield  {title} {\enquote {\bibinfo {title} {{Singlet generation in
					mixed-state quantum networks}},}\ }\href {\doibase
		10.1103/PhysRevA.81.042316} {\bibfield  {journal} {\bibinfo  {journal} {Phys.
				Rev. A}\ }\textbf {\bibinfo {volume} {81}},\ \bibinfo {pages} {042316}
		(\bibinfo {year} {2010})}\BibitemShut {NoStop}%
	\bibitem [{\citenamefont {Bohm}\ and\ \citenamefont
		{Aharonov}(1957)}]{bohm1957}%
	\BibitemOpen
	\bibfield  {author} {\bibinfo {author} {\bibfnamefont {D.}~\bibnamefont
			{Bohm}}\ and\ \bibinfo {author} {\bibfnamefont {Y.}~\bibnamefont
			{Aharonov}},\ }\bibfield  {title} {\enquote {\bibinfo {title} {{Discussion of
					Experimental Proof for the Paradox of Einstein, Rosen, and Podolsky}},}\
	}\href {\doibase 10.1103/PhysRev.108.1070} {\bibfield  {journal} {\bibinfo
			{journal} {Phys. Rev.}\ }\textbf {\bibinfo {volume} {108}},\ \bibinfo {pages}
		{1070--1076} (\bibinfo {year} {1957})}\BibitemShut {NoStop}%
	\bibitem [{\citenamefont {Lidar}\ and\ \citenamefont
		{Brun}(2013)}]{lidar2013quantum}%
	\BibitemOpen
	\bibfield  {author} {\bibinfo {author} {\bibfnamefont {Daniel~A}\
			\bibnamefont {Lidar}}\ and\ \bibinfo {author} {\bibfnamefont {Todd~A}\
			\bibnamefont {Brun}},\ }\href@noop {} {\emph {\bibinfo {title} {Quantum error
				correction}}}\ (\bibinfo  {publisher} {Cambridge university press},\ \bibinfo
	{year} {2013})\BibitemShut {NoStop}%
	\bibitem [{\citenamefont {Devitt}\ \emph {et~al.}(2013)\citenamefont {Devitt},
		\citenamefont {Munro},\ and\ \citenamefont {Nemoto}}]{devitt2013quantum}%
	\BibitemOpen
	\bibfield  {author} {\bibinfo {author} {\bibfnamefont {Simon~J}\ \bibnamefont
			{Devitt}}, \bibinfo {author} {\bibfnamefont {William~J}\ \bibnamefont
			{Munro}}, \ and\ \bibinfo {author} {\bibfnamefont {Kae}\ \bibnamefont
			{Nemoto}},\ }\bibfield  {title} {\enquote {\bibinfo {title} {Quantum error
				correction for beginners},}\ }\href@noop {} {\bibfield  {journal} {\bibinfo
			{journal} {Reports on Progress in Physics}\ }\textbf {\bibinfo {volume}
			{76}},\ \bibinfo {pages} {076001} (\bibinfo {year} {2013})}\BibitemShut
	{NoStop}%
	\bibitem [{\citenamefont {Cabello}(2003)}]{cabello2003supersinglets}%
	\BibitemOpen
	\bibfield  {author} {\bibinfo {author} {\bibfnamefont {Ad{\'a}n}\
			\bibnamefont {Cabello}},\ }\bibfield  {title} {\enquote {\bibinfo {title}
			{Supersinglets},}\ }\href@noop {} {\bibfield  {journal} {\bibinfo  {journal}
			{Journal of Modern Optics}\ }\textbf {\bibinfo {volume} {50}},\ \bibinfo
		{pages} {1049--1061} (\bibinfo {year} {2003})}\BibitemShut {NoStop}%
	\bibitem [{\citenamefont {T{\'o}th}\ and\ \citenamefont
		{Mitchell}(2010)}]{toth2010generation}%
	\BibitemOpen
	\bibfield  {author} {\bibinfo {author} {\bibfnamefont {G{\'e}za}\
			\bibnamefont {T{\'o}th}}\ and\ \bibinfo {author} {\bibfnamefont {Morgan~W}\
			\bibnamefont {Mitchell}},\ }\bibfield  {title} {\enquote {\bibinfo {title}
			{Generation of macroscopic singlet states in atomic ensembles},}\ }\href@noop
	{} {\bibfield  {journal} {\bibinfo  {journal} {New Journal of Physics}\
		}\textbf {\bibinfo {volume} {12}},\ \bibinfo {pages} {053007} (\bibinfo
		{year} {2010})}\BibitemShut {NoStop}%
	\bibitem [{\citenamefont {Horodecki}\ \emph {et~al.}(1999)\citenamefont
		{Horodecki}, \citenamefont {Horodecki},\ and\ \citenamefont
		{Horodecki}}]{horodecki1999general}%
	\BibitemOpen
	\bibfield  {author} {\bibinfo {author} {\bibfnamefont {Micha{\l}}\
			\bibnamefont {Horodecki}}, \bibinfo {author} {\bibfnamefont {Pawe{\l}}\
			\bibnamefont {Horodecki}}, \ and\ \bibinfo {author} {\bibfnamefont {Ryszard}\
			\bibnamefont {Horodecki}},\ }\bibfield  {title} {\enquote {\bibinfo {title}
			{General teleportation channel, singlet fraction, and quasidistillation},}\
	}\href@noop {} {\bibfield  {journal} {\bibinfo  {journal} {Physical Review
				A}\ }\textbf {\bibinfo {volume} {60}},\ \bibinfo {pages} {1888} (\bibinfo
		{year} {1999})}\BibitemShut {NoStop}%
	\bibitem [{\citenamefont {Bennett}\ \emph
		{et~al.}(1996{\natexlab{a}})\citenamefont {Bennett}, \citenamefont
		{Brassard}, \citenamefont {Popescu}, \citenamefont {Schumacher},
		\citenamefont {Smolin},\ and\ \citenamefont
		{Wootters}}]{bennett1996purification}%
	\BibitemOpen
	\bibfield  {author} {\bibinfo {author} {\bibfnamefont {Charles~H}\
			\bibnamefont {Bennett}}, \bibinfo {author} {\bibfnamefont {Gilles}\
			\bibnamefont {Brassard}}, \bibinfo {author} {\bibfnamefont {Sandu}\
			\bibnamefont {Popescu}}, \bibinfo {author} {\bibfnamefont {Benjamin}\
			\bibnamefont {Schumacher}}, \bibinfo {author} {\bibfnamefont {John~A}\
			\bibnamefont {Smolin}}, \ and\ \bibinfo {author} {\bibfnamefont {William~K}\
			\bibnamefont {Wootters}},\ }\bibfield  {title} {\enquote {\bibinfo {title}
			{Purification of noisy entanglement and faithful teleportation via noisy
				channels},}\ }\href@noop {} {\bibfield  {journal} {\bibinfo  {journal}
			{Physical review letters}\ }\textbf {\bibinfo {volume} {76}},\ \bibinfo
		{pages} {722} (\bibinfo {year} {1996}{\natexlab{a}})}\BibitemShut {NoStop}%
	\bibitem [{\citenamefont {Pyrkov}\ and\ \citenamefont
		{Byrnes}(2014)}]{pyrkov2014full}%
	\BibitemOpen
	\bibfield  {author} {\bibinfo {author} {\bibfnamefont {Alexey~N}\
			\bibnamefont {Pyrkov}}\ and\ \bibinfo {author} {\bibfnamefont {Tim}\
			\bibnamefont {Byrnes}},\ }\bibfield  {title} {\enquote {\bibinfo {title}
			{Full-bloch-sphere teleportation of spinor bose-einstein condensates and spin
				ensembles},}\ }\href@noop {} {\bibfield  {journal} {\bibinfo  {journal}
			{Physical Review A}\ }\textbf {\bibinfo {volume} {90}},\ \bibinfo {pages}
		{062336} (\bibinfo {year} {2014})}\BibitemShut {NoStop}%
	\bibitem [{\citenamefont {Byrnes}\ \emph {et~al.}(2012)\citenamefont {Byrnes},
		\citenamefont {Wen},\ and\ \citenamefont {Yamamoto}}]{byrnes2012}%
	\BibitemOpen
	\bibfield  {author} {\bibinfo {author} {\bibfnamefont {T.}~\bibnamefont
			{Byrnes}}, \bibinfo {author} {\bibfnamefont {K.}~\bibnamefont {Wen}}, \ and\
		\bibinfo {author} {\bibfnamefont {Y.}~\bibnamefont {Yamamoto}},\ }\bibfield
	{title} {\enquote {\bibinfo {title} {{Macroscopic quantum computation using
					Bose-Einstein condensates}},}\ }\href@noop {} {\bibfield  {journal} {\bibinfo
			{journal} {Phys. Rev. A}\ }\textbf {\bibinfo {volume} {85}},\ \bibinfo
		{pages} {040306(R)} (\bibinfo {year} {2012})}\BibitemShut {NoStop}%
	\bibitem [{\citenamefont {Abdelrahman}\ \emph {et~al.}(2014)\citenamefont
		{Abdelrahman}, \citenamefont {Mukai}, \citenamefont {H{\"a}ffner},\ and\
		\citenamefont {Byrnes}}]{abdelrahman2014coherent}%
	\BibitemOpen
	\bibfield  {author} {\bibinfo {author} {\bibfnamefont {Ahmed}\ \bibnamefont
			{Abdelrahman}}, \bibinfo {author} {\bibfnamefont {Tetsuya}\ \bibnamefont
			{Mukai}}, \bibinfo {author} {\bibfnamefont {Hartmut}\ \bibnamefont
			{H{\"a}ffner}}, \ and\ \bibinfo {author} {\bibfnamefont {Tim}\ \bibnamefont
			{Byrnes}},\ }\bibfield  {title} {\enquote {\bibinfo {title} {Coherent
				all-optical control of ultracold atoms arrays in permanent magnetic traps},}\
	}\href@noop {} {\bibfield  {journal} {\bibinfo  {journal} {Optics express}\
		}\textbf {\bibinfo {volume} {22}},\ \bibinfo {pages} {3501--3513} (\bibinfo
		{year} {2014})}\BibitemShut {NoStop}%
	\bibitem [{\citenamefont {Bennett}\ \emph
		{et~al.}(1996{\natexlab{b}})\citenamefont {Bennett}, \citenamefont
		{Bernstein}, \citenamefont {Popescu},\ and\ \citenamefont
		{Schumacher}}]{bennett1996c}%
	\BibitemOpen
	\bibfield  {author} {\bibinfo {author} {\bibfnamefont {C.~H.}\ \bibnamefont
			{Bennett}}, \bibinfo {author} {\bibfnamefont {H.~J.}\ \bibnamefont
			{Bernstein}}, \bibinfo {author} {\bibfnamefont {S.}~\bibnamefont {Popescu}},
		\ and\ \bibinfo {author} {\bibfnamefont {B.}~\bibnamefont {Schumacher}},\
	}\bibfield  {title} {\enquote {\bibinfo {title} {Concentrating partial
				entanglement by local operations},}\ }\href@noop {} {\bibfield  {journal}
		{\bibinfo  {journal} {Phys. Rev. A}\ }\textbf {\bibinfo {volume} {53}},\
		\bibinfo {pages} {2046--2052} (\bibinfo {year}
		{1996}{\natexlab{b}})}\BibitemShut {NoStop}%
	\bibitem [{\citenamefont {Behbood}\ \emph {et~al.}(2013)\citenamefont
		{Behbood}, \citenamefont {Colangelo}, \citenamefont {{Martin Ciurana}},
		\citenamefont {Napolitano}, \citenamefont {Sewell},\ and\ \citenamefont
		{Mitchell}}]{behbood2013}%
	\BibitemOpen
	\bibfield  {author} {\bibinfo {author} {\bibfnamefont {N.}~\bibnamefont
			{Behbood}}, \bibinfo {author} {\bibfnamefont {G.}~\bibnamefont {Colangelo}},
		\bibinfo {author} {\bibfnamefont {F.}~\bibnamefont {{Martin Ciurana}}},
		\bibinfo {author} {\bibfnamefont {M.}~\bibnamefont {Napolitano}}, \bibinfo
		{author} {\bibfnamefont {R.~J.}\ \bibnamefont {Sewell}}, \ and\ \bibinfo
		{author} {\bibfnamefont {M.~W.}\ \bibnamefont {Mitchell}},\ }\bibfield
	{title} {\enquote {\bibinfo {title} {{Feedback Cooling of an Atomic Spin
					Ensemble}},}\ }\href {\doibase 10.1103/PhysRevLett.111.103601} {\bibfield
		{journal} {\bibinfo  {journal} {Phys. Rev. Lett.}\ }\textbf {\bibinfo
			{volume} {111}},\ \bibinfo {pages} {103601} (\bibinfo {year}
		{2013})}\BibitemShut {NoStop}%
	\bibitem [{\citenamefont {Behbood}\ \emph {et~al.}(2014)\citenamefont
		{Behbood}, \citenamefont {{Martin Ciurana}}, \citenamefont {Colangelo},
		\citenamefont {Napolitano}, \citenamefont {T{\'o}th}, \citenamefont
		{Sewell},\ and\ \citenamefont {Mitchell}}]{behbood2014}%
	\BibitemOpen
	\bibfield  {author} {\bibinfo {author} {\bibfnamefont {N.}~\bibnamefont
			{Behbood}}, \bibinfo {author} {\bibfnamefont {F.}~\bibnamefont {{Martin
					Ciurana}}}, \bibinfo {author} {\bibfnamefont {G.}~\bibnamefont {Colangelo}},
		\bibinfo {author} {\bibfnamefont {M.}~\bibnamefont {Napolitano}}, \bibinfo
		{author} {\bibfnamefont {G.}~\bibnamefont {T{\'o}th}}, \bibinfo {author}
		{\bibfnamefont {R.~J.}\ \bibnamefont {Sewell}}, \ and\ \bibinfo {author}
		{\bibfnamefont {M.~W.}\ \bibnamefont {Mitchell}},\ }\bibfield  {title}
	{\enquote {\bibinfo {title} {Generation of macroscopic singlet states in a
				cold atomic ensemble},}\ }\href {\doibase 10.1103/PhysRevLett.113.093601}
	{\bibfield  {journal} {\bibinfo  {journal} {Phys. Rev. Lett}\ }\textbf
		{\bibinfo {volume} {113}},\ \bibinfo {pages} {093601} (\bibinfo {year}
		{2014})}\BibitemShut {NoStop}%
	\bibitem [{\citenamefont {Kong}\ \emph {et~al.}(2020)\citenamefont {Kong},
		\citenamefont {Jim{\'e}nez-Mart{\'i}nez}, \citenamefont {Troullinou},
		\citenamefont {Lucivero}, \citenamefont {T{\'o}th},\ and\ \citenamefont
		{Mitchell}}]{kong2020}%
	\BibitemOpen
	\bibfield  {author} {\bibinfo {author} {\bibfnamefont {J.}~\bibnamefont
			{Kong}}, \bibinfo {author} {\bibfnamefont {R.}~\bibnamefont
			{Jim{\'e}nez-Mart{\'i}nez}}, \bibinfo {author} {\bibfnamefont
			{C.}~\bibnamefont {Troullinou}}, \bibinfo {author} {\bibfnamefont {V.~G.}\
			\bibnamefont {Lucivero}}, \bibinfo {author} {\bibfnamefont {G.}~\bibnamefont
			{T{\'o}th}}, \ and\ \bibinfo {author} {\bibfnamefont {M.~W.}\ \bibnamefont
			{Mitchell}},\ }\bibfield  {title} {\enquote {\bibinfo {title}
			{Measurement-induced, spatially-extended entanglement in a hot,
				strongly-interacting atomic system},}\ }\href {\doibase
		doi.org/10.1038/s41467-020-15899-1} {\bibfield  {journal} {\bibinfo
			{journal} {Nat Commun}\ }\textbf {\bibinfo {volume} {11}},\ \bibinfo {pages}
		{2415} (\bibinfo {year} {2020})}\BibitemShut {NoStop}%
	\bibitem [{\citenamefont {Cabello}(2002)}]{cabello2002}%
	\BibitemOpen
	\bibfield  {author} {\bibinfo {author} {\bibfnamefont {A.}~\bibnamefont
			{Cabello}},\ }\bibfield  {title} {\enquote {\bibinfo {title} {{$N$-Particle
					$N$-Level Singlet States: Some Properties and Applications}},}\ }\href
	{\doibase 10.1103/PhysRevLett.89.100402} {\bibfield  {journal} {\bibinfo
			{journal} {Phys. Rev. Lett.}\ }\textbf {\bibinfo {volume} {89}},\ \bibinfo
		{pages} {100402} (\bibinfo {year} {2002})}\BibitemShut {NoStop}%
	\bibitem [{\citenamefont {Jin}\ \emph {et~al.}(2005)\citenamefont {Jin},
		\citenamefont {Li}, \citenamefont {Feng},\ and\ \citenamefont
		{Zheng}}]{jin2005generation}%
	\BibitemOpen
	\bibfield  {author} {\bibinfo {author} {\bibfnamefont {Guang-Sheng}\
			\bibnamefont {Jin}}, \bibinfo {author} {\bibfnamefont {Shu-Shen}\
			\bibnamefont {Li}}, \bibinfo {author} {\bibfnamefont {Song-Lin}\ \bibnamefont
			{Feng}}, \ and\ \bibinfo {author} {\bibfnamefont {Hou-Zhi}\ \bibnamefont
			{Zheng}},\ }\bibfield  {title} {\enquote {\bibinfo {title} {Generation of a
				supersinglet of three three-level atoms in cavity qed},}\ }\href@noop {}
	{\bibfield  {journal} {\bibinfo  {journal} {Physical Review A}\ }\textbf
		{\bibinfo {volume} {71}},\ \bibinfo {pages} {034307} (\bibinfo {year}
		{2005})}\BibitemShut {NoStop}%
	\bibitem [{\citenamefont {Qiang}\ \emph {et~al.}(2011)\citenamefont {Qiang},
		\citenamefont {Cardoso}, \citenamefont {Avelar},\ and\ \citenamefont
		{Baseia}}]{qiang2011alternative}%
	\BibitemOpen
	\bibfield  {author} {\bibinfo {author} {\bibfnamefont {W-C}\ \bibnamefont
			{Qiang}}, \bibinfo {author} {\bibfnamefont {WB}~\bibnamefont {Cardoso}},
		\bibinfo {author} {\bibfnamefont {AT}~\bibnamefont {Avelar}}, \ and\ \bibinfo
		{author} {\bibfnamefont {B}~\bibnamefont {Baseia}},\ }\bibfield  {title}
	{\enquote {\bibinfo {title} {Alternative scheme to generate a supersinglet
				state of three-level atoms},}\ }\href@noop {} {\bibfield  {journal} {\bibinfo
			{journal} {Physics Letters A}\ }\textbf {\bibinfo {volume} {375}},\ \bibinfo
		{pages} {443--447} (\bibinfo {year} {2011})}\BibitemShut {NoStop}%
	\bibitem [{\citenamefont {Chen}\ \emph {et~al.}(2016)\citenamefont {Chen},
		\citenamefont {Chen}, \citenamefont {Xia}, \citenamefont {Song},\ and\
		\citenamefont {Huang}}]{chen2016fast}%
	\BibitemOpen
	\bibfield  {author} {\bibinfo {author} {\bibfnamefont {Zhen}\ \bibnamefont
			{Chen}}, \bibinfo {author} {\bibfnamefont {Ye-Hong}\ \bibnamefont {Chen}},
		\bibinfo {author} {\bibfnamefont {Yan}\ \bibnamefont {Xia}}, \bibinfo
		{author} {\bibfnamefont {Jie}\ \bibnamefont {Song}}, \ and\ \bibinfo {author}
		{\bibfnamefont {Bi-Hua}\ \bibnamefont {Huang}},\ }\bibfield  {title}
	{\enquote {\bibinfo {title} {Fast generation of three-atom singlet state by
				transitionless quantum driving},}\ }\href@noop {} {\bibfield  {journal}
		{\bibinfo  {journal} {Scientific Reports}\ }\textbf {\bibinfo {volume} {6}},\
		\bibinfo {pages} {1--12} (\bibinfo {year} {2016})}\BibitemShut {NoStop}%
	\bibitem [{\citenamefont {M.~Schimpf}\ and\ \citenamefont
		{Svozil}(2010)}]{schimpf2010}%
	\BibitemOpen
	\bibfield  {author} {\bibinfo {author} {\bibfnamefont {Maria}\ \bibnamefont
			{M.~Schimpf}}\ and\ \bibinfo {author} {\bibfnamefont {K.}~\bibnamefont
			{Svozil}},\ }\bibfield  {title} {\enquote {\bibinfo {title} {{A glance at
					singlet states and four-partite correlations}},}\ }\href@noop {} {\bibfield
		{journal} {\bibinfo  {journal} {Mathematica Slovaca}\ }\textbf {\bibinfo
			{volume} {60}},\ \bibinfo {pages} {701--722} (\bibinfo {year}
		{2010})}\BibitemShut {NoStop}%
	\bibitem [{\citenamefont {B{\"o}hi}\ \emph {et~al.}(2009)\citenamefont
		{B{\"o}hi}, \citenamefont {Riedel}, \citenamefont {Hoffrogge}, \citenamefont
		{Reichel}, \citenamefont {H{\"a}nsch},\ and\ \citenamefont
		{P.Treutlein}}]{bohi2009}%
	\BibitemOpen
	\bibfield  {author} {\bibinfo {author} {\bibfnamefont {P.}~\bibnamefont
			{B{\"o}hi}}, \bibinfo {author} {\bibfnamefont {M.~F.}\ \bibnamefont
			{Riedel}}, \bibinfo {author} {\bibfnamefont {J.}~\bibnamefont {Hoffrogge}},
		\bibinfo {author} {\bibfnamefont {J.}~\bibnamefont {Reichel}}, \bibinfo
		{author} {\bibfnamefont {T.~W.}\ \bibnamefont {H{\"a}nsch}}, \ and\ \bibinfo
		{author} {\bibnamefont {P.Treutlein}},\ }\bibfield  {title} {\enquote
		{\bibinfo {title} {Coherent manipulation of {Bose–Einstein} condensates
				with state-dependent microwave potentials on an atom chip},}\ }\href@noop {}
	{\bibfield  {journal} {\bibinfo  {journal} {Nature Physics}\ }\textbf
		{\bibinfo {volume} {5}},\ \bibinfo {pages} {592} (\bibinfo {year}
		{2009})}\BibitemShut {NoStop}%
	\bibitem [{\citenamefont {Riedel}\ \emph {et~al.}(2010)\citenamefont {Riedel},
		\citenamefont {B{\"o}hi}, \citenamefont {Li}, \citenamefont {H{\"a}nsch},
		\citenamefont {Sinatra},\ and\ \citenamefont {Treutlein}}]{riedel2010}%
	\BibitemOpen
	\bibfield  {author} {\bibinfo {author} {\bibfnamefont {M.~F.}\ \bibnamefont
			{Riedel}}, \bibinfo {author} {\bibfnamefont {P.}~\bibnamefont {B{\"o}hi}},
		\bibinfo {author} {\bibfnamefont {Y.}~\bibnamefont {Li}}, \bibinfo {author}
		{\bibfnamefont {T.~W.}\ \bibnamefont {H{\"a}nsch}}, \bibinfo {author}
		{\bibfnamefont {A.}~\bibnamefont {Sinatra}}, \ and\ \bibinfo {author}
		{\bibfnamefont {P.}~\bibnamefont {Treutlein}},\ }\bibfield  {title} {\enquote
		{\bibinfo {title} {Atom-chip-based generation of entanglement for quantum
				metrology},}\ }\href@noop {} {\bibfield  {journal} {\bibinfo  {journal}
			{Nature}\ }\textbf {\bibinfo {volume} {464}},\ \bibinfo {pages} {1170}
		(\bibinfo {year} {2010})}\BibitemShut {NoStop}%
	\bibitem [{\citenamefont {Hammerer}\ \emph {et~al.}(2010)\citenamefont
		{Hammerer}, \citenamefont {S\o{}rensen},\ and\ \citenamefont
		{Polzik}}]{hammerer2010}%
	\BibitemOpen
	\bibfield  {author} {\bibinfo {author} {\bibfnamefont {K.}~\bibnamefont
			{Hammerer}}, \bibinfo {author} {\bibfnamefont {A.~S.}\ \bibnamefont
			{S\o{}rensen}}, \ and\ \bibinfo {author} {\bibfnamefont {E.~S.}\ \bibnamefont
			{Polzik}},\ }\bibfield  {title} {\enquote {\bibinfo {title} {{Quantum
					interface between light and atomic ensembles}},}\ }\href {\doibase
		10.1103/RevModPhys.82.1041} {\bibfield  {journal} {\bibinfo  {journal} {Rev.
				Mod. Phys.}\ }\textbf {\bibinfo {volume} {82}},\ \bibinfo {pages}
		{1041--1093} (\bibinfo {year} {2010})}\BibitemShut {NoStop}%
	\bibitem [{\citenamefont {Pezz{\`e}}\ \emph {et~al.}(2018)\citenamefont
		{Pezz{\`e}}, \citenamefont {Smerzi}, \citenamefont {Oberthaler},
		\citenamefont {Schmied},\ and\ \citenamefont {Treutlein}}]{pezze2018}%
	\BibitemOpen
	\bibfield  {author} {\bibinfo {author} {\bibfnamefont {L.}~\bibnamefont
			{Pezz{\`e}}}, \bibinfo {author} {\bibfnamefont {A.}~\bibnamefont {Smerzi}},
		\bibinfo {author} {\bibfnamefont {M.~K.}\ \bibnamefont {Oberthaler}},
		\bibinfo {author} {\bibfnamefont {R.}~\bibnamefont {Schmied}}, \ and\
		\bibinfo {author} {\bibfnamefont {P.}~\bibnamefont {Treutlein}},\ }\bibfield
	{title} {\enquote {\bibinfo {title} {Quantum metrology with nonclassical
				states of atomic ensembles},}\ }\href {\doibase 10.1103/RevModPhys.90.035005}
	{\bibfield  {journal} {\bibinfo  {journal} {Rev. Mod. Phys.}\ }\textbf
		{\bibinfo {volume} {90}},\ \bibinfo {pages} {035005} (\bibinfo {year}
		{2018})}\BibitemShut {NoStop}%
	\bibitem [{\citenamefont {Takahashi}\ \emph {et~al.}(1999)\citenamefont
		{Takahashi}, \citenamefont {Honda}, \citenamefont {Tanaka}, \citenamefont
		{Toyoda}, \citenamefont {Ishikawa},\ and\ \citenamefont
		{Yabuzaki}}]{takahashi1999}%
	\BibitemOpen
	\bibfield  {author} {\bibinfo {author} {\bibfnamefont {Y.}~\bibnamefont
			{Takahashi}}, \bibinfo {author} {\bibfnamefont {K.}~\bibnamefont {Honda}},
		\bibinfo {author} {\bibfnamefont {N.}~\bibnamefont {Tanaka}}, \bibinfo
		{author} {\bibfnamefont {K.}~\bibnamefont {Toyoda}}, \bibinfo {author}
		{\bibfnamefont {K.}~\bibnamefont {Ishikawa}}, \ and\ \bibinfo {author}
		{\bibfnamefont {T.}~\bibnamefont {Yabuzaki}},\ }\bibfield  {title} {\enquote
		{\bibinfo {title} {Quantum nondemolition measurement of spin via the
				paramagnetic faraday rotation},}\ }\href@noop {} {\bibfield  {journal}
		{\bibinfo  {journal} {Phys. Rev. A}\ }\textbf {\bibinfo {volume} {60}},\
		\bibinfo {pages} {4974--4979} (\bibinfo {year} {1999})}\BibitemShut {NoStop}%
	\bibitem [{\citenamefont {Higbie}\ \emph {et~al.}(2005)\citenamefont {Higbie},
		\citenamefont {Sadler}, \citenamefont {Inouye}, \citenamefont {Chikkatur},
		\citenamefont {Leslie}, \citenamefont {Moore}, \citenamefont {Savalli},\ and\
		\citenamefont {Stamper-Kurn}}]{higbie2005}%
	\BibitemOpen
	\bibfield  {author} {\bibinfo {author} {\bibfnamefont {J.~M.}\ \bibnamefont
			{Higbie}}, \bibinfo {author} {\bibfnamefont {L.~E.}\ \bibnamefont {Sadler}},
		\bibinfo {author} {\bibfnamefont {S.}~\bibnamefont {Inouye}}, \bibinfo
		{author} {\bibfnamefont {A.~P.}\ \bibnamefont {Chikkatur}}, \bibinfo {author}
		{\bibfnamefont {S.~R.}\ \bibnamefont {Leslie}}, \bibinfo {author}
		{\bibfnamefont {K.~L.}\ \bibnamefont {Moore}}, \bibinfo {author}
		{\bibfnamefont {V.}~\bibnamefont {Savalli}}, \ and\ \bibinfo {author}
		{\bibfnamefont {D.~M.}\ \bibnamefont {Stamper-Kurn}},\ }\bibfield  {title}
	{\enquote {\bibinfo {title} {{Direct Nondestructive Imaging of Magnetization
					in a Spin-1 Bose-Einstein Gas}},}\ }\href@noop {} {\bibfield  {journal}
		{\bibinfo  {journal} {Phys. Rev. Lett.}\ }\textbf {\bibinfo {volume} {95}},\
		\bibinfo {pages} {050401} (\bibinfo {year} {2005})}\BibitemShut {NoStop}%
	\bibitem [{\citenamefont {Kuzmich}\ and\ \citenamefont
		{Kennedy}(2004)}]{kuzmich2004}%
	\BibitemOpen
	\bibfield  {author} {\bibinfo {author} {\bibfnamefont {A.}~\bibnamefont
			{Kuzmich}}\ and\ \bibinfo {author} {\bibfnamefont {T.~A.~B.}\ \bibnamefont
			{Kennedy}},\ }\bibfield  {title} {\enquote {\bibinfo {title} {Nonsymmetric
				entanglement of atomic ensembles},}\ }\href {\doibase
		10.1103/PhysRevLett.92.030407} {\bibfield  {journal} {\bibinfo  {journal}
			{Phys. Rev. Lett.}\ }\textbf {\bibinfo {volume} {92}},\ \bibinfo {pages}
		{030407} (\bibinfo {year} {2004})}\BibitemShut {NoStop}%
	\bibitem [{\citenamefont {Meppelink}\ \emph {et~al.}(2010)\citenamefont
		{Meppelink}, \citenamefont {Rozendaal}, \citenamefont {Koller}, \citenamefont
		{Vogels},\ and\ \citenamefont {van~der Straten}}]{meppelink2010}%
	\BibitemOpen
	\bibfield  {author} {\bibinfo {author} {\bibfnamefont {R.}~\bibnamefont
			{Meppelink}}, \bibinfo {author} {\bibfnamefont {R.~A.}\ \bibnamefont
			{Rozendaal}}, \bibinfo {author} {\bibfnamefont {S.~B.}\ \bibnamefont
			{Koller}}, \bibinfo {author} {\bibfnamefont {J.~M.}\ \bibnamefont {Vogels}},
		\ and\ \bibinfo {author} {\bibfnamefont {P.}~\bibnamefont {van~der
				Straten}},\ }\bibfield  {title} {\enquote {\bibinfo {title} {{Thermodynamics
					of Bose-Einstein-condensed clouds using phase-contrast imaging}},}\
	}\href@noop {} {\bibfield  {journal} {\bibinfo  {journal} {Phys. Rev. A}\
		}\textbf {\bibinfo {volume} {81}},\ \bibinfo {pages} {053632} (\bibinfo
		{year} {2010})}\BibitemShut {NoStop}%
	\bibitem [{\citenamefont {Ilo-Okeke}\ and\ \citenamefont
		{Byrnes}(2014)}]{ilo-okeke2014}%
	\BibitemOpen
	\bibfield  {author} {\bibinfo {author} {\bibfnamefont {E.~O.}\ \bibnamefont
			{Ilo-Okeke}}\ and\ \bibinfo {author} {\bibfnamefont {T.}~\bibnamefont
			{Byrnes}},\ }\bibfield  {title} {\enquote {\bibinfo {title} {Theory of
				single-shot phase contrast imaging in spinor {Bose-Einstein} condensates},}\
	}\href@noop {} {\bibfield  {journal} {\bibinfo  {journal} {Phys. Rev. Lett.}\
		}\textbf {\bibinfo {volume} {112}},\ \bibinfo {pages} {233602} (\bibinfo
		{year} {2014})}\BibitemShut {NoStop}%
	\bibitem [{\citenamefont {Ilo-Okeke}\ and\ \citenamefont
		{Byrnes}(2016)}]{ilo-okeke2016}%
	\BibitemOpen
	\bibfield  {author} {\bibinfo {author} {\bibfnamefont {E.~O.}\ \bibnamefont
			{Ilo-Okeke}}\ and\ \bibinfo {author} {\bibfnamefont {T.}~\bibnamefont
			{Byrnes}},\ }\bibfield  {title} {\enquote {\bibinfo {title} {Information and
				backaction due to phase-contrast-imaging measurements of cold atomic gases:
				Beyond gaussian states},}\ }\href@noop {} {\bibfield  {journal} {\bibinfo
			{journal} {Phys. Rev. A}\ }\textbf {\bibinfo {volume} {94}},\ \bibinfo
		{pages} {013617} (\bibinfo {year} {2016})}\BibitemShut {NoStop}%
	\bibitem [{\citenamefont {Appel}\ \emph {et~al.}(2009)\citenamefont {Appel},
		\citenamefont {Windpassinger}, \citenamefont {Oblak}, \citenamefont {Hoff},
		\citenamefont {Kj{\ae}rgaard},\ and\ \citenamefont {Polzik}}]{appel2009}%
	\BibitemOpen
	\bibfield  {author} {\bibinfo {author} {\bibfnamefont {J.}~\bibnamefont
			{Appel}}, \bibinfo {author} {\bibfnamefont {P.~J.}\ \bibnamefont
			{Windpassinger}}, \bibinfo {author} {\bibfnamefont {D.}~\bibnamefont
			{Oblak}}, \bibinfo {author} {\bibfnamefont {U.~B.}\ \bibnamefont {Hoff}},
		\bibinfo {author} {\bibfnamefont {N.}~\bibnamefont {Kj{\ae}rgaard}}, \ and\
		\bibinfo {author} {\bibfnamefont {E.~S.}\ \bibnamefont {Polzik}},\ }\bibfield
	{title} {\enquote {\bibinfo {title} {Mesoscopic atomic entanglement for
				precision measurements beyond the standard quantum limit},}\ }\href@noop {}
	{\bibfield  {journal} {\bibinfo  {journal} {Proc. Natl. Acad. Sci. USA}\
		}\textbf {\bibinfo {volume} {106}},\ \bibinfo {pages} {10960--10965}
		(\bibinfo {year} {2009})}\BibitemShut {NoStop}%
	\bibitem [{\citenamefont {Schleier-Smith}\ \emph {et~al.}(2010)\citenamefont
		{Schleier-Smith}, \citenamefont {Leroux},\ and\ \citenamefont
		{Vuleti{\'c}}}]{schleier-smith2010}%
	\BibitemOpen
	\bibfield  {author} {\bibinfo {author} {\bibfnamefont {M.~H.}\ \bibnamefont
			{Schleier-Smith}}, \bibinfo {author} {\bibfnamefont {I.~D.}\ \bibnamefont
			{Leroux}}, \ and\ \bibinfo {author} {\bibfnamefont {V.}~\bibnamefont
			{Vuleti{\'c}}},\ }\bibfield  {title} {\enquote {\bibinfo {title} {States of
				an ensemble of two-level atoms with reduced quantum uncertainty},}\
	}\href@noop {} {\bibfield  {journal} {\bibinfo  {journal} {Phys. Rev. Lett.}\
		}\textbf {\bibinfo {volume} {104}},\ \bibinfo {pages} {073604} (\bibinfo
		{year} {2010})}\BibitemShut {NoStop}%
	\bibitem [{\citenamefont {Sewell}\ \emph {et~al.}(2012)\citenamefont {Sewell},
		\citenamefont {Koschorreck}, \citenamefont {Napolitano}, \citenamefont
		{Dubost}, \citenamefont {Behbood},\ and\ \citenamefont
		{Mitchell}}]{sewell2012}%
	\BibitemOpen
	\bibfield  {author} {\bibinfo {author} {\bibfnamefont {R.~J.}\ \bibnamefont
			{Sewell}}, \bibinfo {author} {\bibfnamefont {M.}~\bibnamefont {Koschorreck}},
		\bibinfo {author} {\bibfnamefont {M.}~\bibnamefont {Napolitano}}, \bibinfo
		{author} {\bibfnamefont {B.}~\bibnamefont {Dubost}}, \bibinfo {author}
		{\bibfnamefont {N.}~\bibnamefont {Behbood}}, \ and\ \bibinfo {author}
		{\bibfnamefont {M.~W.}\ \bibnamefont {Mitchell}},\ }\bibfield  {title}
	{\enquote {\bibinfo {title} {Magnetic sensitivity beyond the projection noise
				limit by spin squeezing},}\ }\href@noop {} {\bibfield  {journal} {\bibinfo
			{journal} {Phys. Rev. Lett.}\ }\textbf {\bibinfo {volume} {109}},\ \bibinfo
		{pages} {253605} (\bibinfo {year} {2012})}\BibitemShut {NoStop}%
	\bibitem [{\citenamefont {Cox}\ \emph {et~al.}(2016)\citenamefont {Cox},
		\citenamefont {Greve}, \citenamefont {Weiner},\ and\ \citenamefont
		{Thompson}}]{cox2016}%
	\BibitemOpen
	\bibfield  {author} {\bibinfo {author} {\bibfnamefont {K.~C.}\ \bibnamefont
			{Cox}}, \bibinfo {author} {\bibfnamefont {G.~P.}\ \bibnamefont {Greve}},
		\bibinfo {author} {\bibfnamefont {J.~M.}\ \bibnamefont {Weiner}}, \ and\
		\bibinfo {author} {\bibfnamefont {J.~K.}\ \bibnamefont {Thompson}},\
	}\bibfield  {title} {\enquote {\bibinfo {title} {Deterministic squeezed
				states with collective measurements and feedback},}\ }\href {\doibase
		10.1103/PhysRevLett.116.093602} {\bibfield  {journal} {\bibinfo  {journal}
			{Phys. Rev. Lett.}\ }\textbf {\bibinfo {volume} {116}},\ \bibinfo {pages}
		{093602} (\bibinfo {year} {2016})}\BibitemShut {NoStop}%
	\bibitem [{\citenamefont {Hosten}\ \emph {et~al.}(2016)\citenamefont {Hosten},
		\citenamefont {Engelsen}, \citenamefont {Krishnakumar},\ and\ \citenamefont
		{Kasevich}}]{hosten2016}%
	\BibitemOpen
	\bibfield  {author} {\bibinfo {author} {\bibfnamefont {O.}~\bibnamefont
			{Hosten}}, \bibinfo {author} {\bibfnamefont {N.~J.}\ \bibnamefont
			{Engelsen}}, \bibinfo {author} {\bibfnamefont {R.}~\bibnamefont
			{Krishnakumar}}, \ and\ \bibinfo {author} {\bibfnamefont {M.~A.}\
			\bibnamefont {Kasevich}},\ }\bibfield  {title} {\enquote {\bibinfo {title}
			{Measurement noise 100 times lower than the quantum-projection limit using
				entangled atoms},}\ }\href {\doibase 10.1038/nature16176} {\bibfield
		{journal} {\bibinfo  {journal} {Nature}\ }\textbf {\bibinfo {volume} {529}},\
		\bibinfo {pages} {505--508} (\bibinfo {year} {2016})}\BibitemShut {NoStop}%
	\bibitem [{\citenamefont {Vasilakis}\ \emph {et~al.}(2015)\citenamefont
		{Vasilakis}, \citenamefont {Shen}, \citenamefont {Jensen}, \citenamefont
		{Balabas}, \citenamefont {Salart}, \citenamefont {Chen},\ and\ \citenamefont
		{Polzik}}]{vasilakis2015}%
	\BibitemOpen
	\bibfield  {author} {\bibinfo {author} {\bibfnamefont {G.}~\bibnamefont
			{Vasilakis}}, \bibinfo {author} {\bibfnamefont {H.}~\bibnamefont {Shen}},
		\bibinfo {author} {\bibfnamefont {K.}~\bibnamefont {Jensen}}, \bibinfo
		{author} {\bibfnamefont {M.}~\bibnamefont {Balabas}}, \bibinfo {author}
		{\bibfnamefont {D.}~\bibnamefont {Salart}}, \bibinfo {author} {\bibfnamefont
			{B.}~\bibnamefont {Chen}}, \ and\ \bibinfo {author} {\bibfnamefont {E.~S.}\
			\bibnamefont {Polzik}},\ }\bibfield  {title} {\enquote {\bibinfo {title}
			{{Generation of a squeezed state of an oscillator by stroboscopic
					back-action-evading measurement}},}\ }\href@noop {} {\bibfield  {journal}
		{\bibinfo  {journal} {Nature Physics}\ }\textbf {\bibinfo {volume} {11}},\
		\bibinfo {pages} {389--392} (\bibinfo {year} {2015})}\BibitemShut {NoStop}%
	\bibitem [{\citenamefont {M{\o}ller}\ \emph {et~al.}(2017)\citenamefont
		{M{\o}ller}, \citenamefont {Thomas}, \citenamefont {Vasilakis}, \citenamefont
		{Zeuthen}, \citenamefont {Tsaturyan}, \citenamefont {Balabas}, \citenamefont
		{Jensen}, \citenamefont {Schliesser}, \citenamefont {Hammerer},\ and\
		\citenamefont {Polzik}}]{moller2017}%
	\BibitemOpen
	\bibfield  {author} {\bibinfo {author} {\bibfnamefont {C.~B.}\ \bibnamefont
			{M{\o}ller}}, \bibinfo {author} {\bibfnamefont {R.~A.}\ \bibnamefont
			{Thomas}}, \bibinfo {author} {\bibfnamefont {G.}~\bibnamefont {Vasilakis}},
		\bibinfo {author} {\bibfnamefont {E.}~\bibnamefont {Zeuthen}}, \bibinfo
		{author} {\bibfnamefont {Y.}~\bibnamefont {Tsaturyan}}, \bibinfo {author}
		{\bibfnamefont {M.}~\bibnamefont {Balabas}}, \bibinfo {author} {\bibfnamefont
			{K.}~\bibnamefont {Jensen}}, \bibinfo {author} {\bibfnamefont
			{A.}~\bibnamefont {Schliesser}}, \bibinfo {author} {\bibfnamefont
			{K.}~\bibnamefont {Hammerer}}, \ and\ \bibinfo {author} {\bibfnamefont
			{E.~S.}\ \bibnamefont {Polzik}},\ }\bibfield  {title} {\enquote {\bibinfo
			{title} {Quantum back-action-evading measurement of motion in a negative mass
				reference frame},}\ }\href@noop {} {\bibfield  {journal} {\bibinfo  {journal}
			{Nature}\ }\textbf {\bibinfo {volume} {547}},\ \bibinfo {pages} {191--195}
		(\bibinfo {year} {2017})}\BibitemShut {NoStop}%
	\bibitem [{\citenamefont {Aristizabal-Zuluaga}\ \emph
		{et~al.}(2021)\citenamefont {Aristizabal-Zuluaga}, \citenamefont {Skobleva},
		\citenamefont {Richter}, \citenamefont {Ji}, \citenamefont {Mao},
		\citenamefont {Kondappan}, \citenamefont {Ivannikov},\ and\ \citenamefont
		{Byrnes}}]{aristizabal-zuluga2021}%
	\BibitemOpen
	\bibfield  {author} {\bibinfo {author} {\bibfnamefont {J.~E.}\ \bibnamefont
			{Aristizabal-Zuluaga}}, \bibinfo {author} {\bibfnamefont {I.}~\bibnamefont
			{Skobleva}}, \bibinfo {author} {\bibfnamefont {L.}~\bibnamefont {Richter}},
		\bibinfo {author} {\bibfnamefont {Y.}~\bibnamefont {Ji}}, \bibinfo {author}
		{\bibfnamefont {Y.}~\bibnamefont {Mao}}, \bibinfo {author} {\bibfnamefont
			{M.}~\bibnamefont {Kondappan}}, \bibinfo {author} {\bibfnamefont
			{V.}~\bibnamefont {Ivannikov}}, \ and\ \bibinfo {author} {\bibfnamefont
			{T.}~\bibnamefont {Byrnes}},\ }\bibfield  {title} {\enquote {\bibinfo {title}
			{{Faraday-imaging-induced squeezing of a double-well Bose-Einstein
					condensate}},}\ }\href {\doibase 10.1088/1361-6455/abf6b5} {\bibfield
		{journal} {\bibinfo  {journal} {J. Phys. B: At. Mol. Opt. Phys.}\ }\textbf
		{\bibinfo {volume} {54}},\ \bibinfo {pages} {105502} (\bibinfo {year}
		{2021})}\BibitemShut {NoStop}%
	\bibitem [{\citenamefont {Ilo-Okeke}\ \emph {et~al.}(2021)\citenamefont
		{Ilo-Okeke}, \citenamefont {Sunami}, \citenamefont {Foot},\ and\
		\citenamefont {Byrnes}}]{ilo-okeke2021}%
	\BibitemOpen
	\bibfield  {author} {\bibinfo {author} {\bibfnamefont {E.~O.}\ \bibnamefont
			{Ilo-Okeke}}, \bibinfo {author} {\bibfnamefont {S.}~\bibnamefont {Sunami}},
		\bibinfo {author} {\bibfnamefont {C.~J.}\ \bibnamefont {Foot}}, \ and\
		\bibinfo {author} {\bibfnamefont {T.}~\bibnamefont {Byrnes}},\ }\bibfield
	{title} {\enquote {\bibinfo {title} {{Faraday-imaging-induced squeezing of a
					double-well Bose-Einstein condensate}},}\ }\href {\doibase
		10.1103/PhysRevA.104.053324} {\bibfield  {journal} {\bibinfo  {journal}
			{Phys. Rev. A}\ }\textbf {\bibinfo {volume} {104}},\ \bibinfo {pages}
		{053324} (\bibinfo {year} {2021})}\BibitemShut {NoStop}%
	\bibitem [{\citenamefont {Kuzmich}\ \emph {et~al.}(2000)\citenamefont
		{Kuzmich}, \citenamefont {Mandel},\ and\ \citenamefont
		{Bigelow}}]{kuzmich2000}%
	\BibitemOpen
	\bibfield  {author} {\bibinfo {author} {\bibfnamefont {A.}~\bibnamefont
			{Kuzmich}}, \bibinfo {author} {\bibfnamefont {L.}~\bibnamefont {Mandel}}, \
		and\ \bibinfo {author} {\bibfnamefont {N.~P.}\ \bibnamefont {Bigelow}},\
	}\bibfield  {title} {\enquote {\bibinfo {title} {Generation of spin squeezing
				via continuous quantum nondemolition measurement},}\ }\href {\doibase
		10.1103/PhysRevLett.85.1594} {\bibfield  {journal} {\bibinfo  {journal}
			{Phys. Rev. Lett.}\ }\textbf {\bibinfo {volume} {85}},\ \bibinfo {pages}
		{1594--1597} (\bibinfo {year} {2000})}\BibitemShut {NoStop}%
	\bibitem [{\citenamefont {Ilo-Okeke}\ and\ \citenamefont
		{Byrnes}(2022)}]{ilo-okeke2022}%
	\BibitemOpen
	\bibfield  {author} {\bibinfo {author} {\bibfnamefont {E.~O.}\ \bibnamefont
			{Ilo-Okeke}}\ and\ \bibinfo {author} {\bibfnamefont {T.}~\bibnamefont
			{Byrnes}},\ }\href@noop {} {\enquote {\bibinfo {title} {Measurement operator
				formalism for quantum nondemolition measurements},}\ } (\bibinfo {year}
	{2022}),\ \bibinfo {note} {in preparation}\BibitemShut {NoStop}%
	\bibitem [{\citenamefont {Genov}\ \emph {et~al.}(2014)\citenamefont {Genov},
		\citenamefont {Schraft}, \citenamefont {Halfmann},\ and\ \citenamefont
		{Vitanov}}]{genov2014correction}%
	\BibitemOpen
	\bibfield  {author} {\bibinfo {author} {\bibfnamefont {Genko~T}\ \bibnamefont
			{Genov}}, \bibinfo {author} {\bibfnamefont {Daniel}\ \bibnamefont {Schraft}},
		\bibinfo {author} {\bibfnamefont {Thomas}\ \bibnamefont {Halfmann}}, \ and\
		\bibinfo {author} {\bibfnamefont {Nikolay~V}\ \bibnamefont {Vitanov}},\
	}\bibfield  {title} {\enquote {\bibinfo {title} {Correction of arbitrary
				field errors in population inversion of quantum systems by universal
				composite pulses},}\ }\href@noop {} {\bibfield  {journal} {\bibinfo
			{journal} {Physical Review Letters}\ }\textbf {\bibinfo {volume} {113}},\
		\bibinfo {pages} {043001} (\bibinfo {year} {2014})}\BibitemShut {NoStop}%
	\bibitem [{\citenamefont {Mao}\ \emph {et~al.}(2022)\citenamefont {Mao},
		\citenamefont {Chaudhary}, \citenamefont {Kondappan}, \citenamefont {Shi},
		\citenamefont {Ilo-Okeke}, \citenamefont {Ivannikov},\ and\ \citenamefont
		{Byrnes}}]{mao2022measurement}%
	\BibitemOpen
	\bibfield  {author} {\bibinfo {author} {\bibfnamefont {Yuping}\ \bibnamefont
			{Mao}}, \bibinfo {author} {\bibfnamefont {Manish}\ \bibnamefont {Chaudhary}},
		\bibinfo {author} {\bibfnamefont {Manikandan}\ \bibnamefont {Kondappan}},
		\bibinfo {author} {\bibfnamefont {Junheng}\ \bibnamefont {Shi}}, \bibinfo
		{author} {\bibfnamefont {Ebubechukwu~O}\ \bibnamefont {Ilo-Okeke}}, \bibinfo
		{author} {\bibfnamefont {Valentin}\ \bibnamefont {Ivannikov}}, \ and\
		\bibinfo {author} {\bibfnamefont {Tim}\ \bibnamefont {Byrnes}},\ }\bibfield
	{title} {\enquote {\bibinfo {title} {Measurement-based deterministic
				imaginary time evolution},}\ }\href@noop {} {\bibfield  {journal} {\bibinfo
			{journal} {arXiv preprint arXiv:2202.09100}\ } (\bibinfo {year}
		{2022})}\BibitemShut {NoStop}%
	\bibitem [{\citenamefont {Radhakrishnan}\ \emph {et~al.}(2019)\citenamefont
		{Radhakrishnan}, \citenamefont {Ding}, \citenamefont {Shi}, \citenamefont
		{Du},\ and\ \citenamefont {Byrnes}}]{radhakrishnan2019basis}%
	\BibitemOpen
	\bibfield  {author} {\bibinfo {author} {\bibfnamefont {Chandrashekar}\
			\bibnamefont {Radhakrishnan}}, \bibinfo {author} {\bibfnamefont {Zhe}\
			\bibnamefont {Ding}}, \bibinfo {author} {\bibfnamefont {Fazhan}\ \bibnamefont
			{Shi}}, \bibinfo {author} {\bibfnamefont {Jiangfeng}\ \bibnamefont {Du}}, \
		and\ \bibinfo {author} {\bibfnamefont {Tim}\ \bibnamefont {Byrnes}},\
	}\bibfield  {title} {\enquote {\bibinfo {title} {Basis-independent quantum
				coherence and its distribution},}\ }\href@noop {} {\bibfield  {journal}
		{\bibinfo  {journal} {Annals of Physics}\ }\textbf {\bibinfo {volume}
			{409}},\ \bibinfo {pages} {167906} (\bibinfo {year} {2019})}\BibitemShut
	{NoStop}%
	\bibitem [{\citenamefont {Ma}\ \emph {et~al.}(2019)\citenamefont {Ma},
		\citenamefont {Cui}, \citenamefont {Cao}, \citenamefont {Fei}, \citenamefont
		{Vedral}, \citenamefont {Byrnes},\ and\ \citenamefont
		{Radhakrishnan}}]{ma2019operational}%
	\BibitemOpen
	\bibfield  {author} {\bibinfo {author} {\bibfnamefont {Zhi-Hao}\ \bibnamefont
			{Ma}}, \bibinfo {author} {\bibfnamefont {Jian}\ \bibnamefont {Cui}}, \bibinfo
		{author} {\bibfnamefont {Zhu}\ \bibnamefont {Cao}}, \bibinfo {author}
		{\bibfnamefont {Shao-Ming}\ \bibnamefont {Fei}}, \bibinfo {author}
		{\bibfnamefont {Vlatko}\ \bibnamefont {Vedral}}, \bibinfo {author}
		{\bibfnamefont {Tim}\ \bibnamefont {Byrnes}}, \ and\ \bibinfo {author}
		{\bibfnamefont {Chandrashekar}\ \bibnamefont {Radhakrishnan}},\ }\bibfield
	{title} {\enquote {\bibinfo {title} {Operational advantage of
				basis-independent quantum coherence},}\ }\href@noop {} {\bibfield  {journal}
		{\bibinfo  {journal} {EPL (Europhysics Letters)}\ }\textbf {\bibinfo {volume}
			{125}},\ \bibinfo {pages} {50005} (\bibinfo {year} {2019})}\BibitemShut
	{NoStop}%
	\bibitem [{\citenamefont {Arecchi}\ \emph {et~al.}(1972)\citenamefont
		{Arecchi}, \citenamefont {Courtens}, \citenamefont {Gilmore},\ and\
		\citenamefont {Thomas}}]{arecchi1972}%
	\BibitemOpen
	\bibfield  {author} {\bibinfo {author} {\bibfnamefont {F.~T.}\ \bibnamefont
			{Arecchi}}, \bibinfo {author} {\bibfnamefont {E.}~\bibnamefont {Courtens}},
		\bibinfo {author} {\bibfnamefont {R.}~\bibnamefont {Gilmore}}, \ and\
		\bibinfo {author} {\bibfnamefont {H.}~\bibnamefont {Thomas}},\ }\bibfield
	{title} {\enquote {\bibinfo {title} {Atomic coherent states in quantum
				optics},}\ }\href@noop {} {\bibfield  {journal} {\bibinfo  {journal} {Phys.
				Rev. A}\ }\textbf {\bibinfo {volume} {6}},\ \bibinfo {pages} {2211--2237}
		(\bibinfo {year} {1972})}\BibitemShut {NoStop}%
\end{thebibliography}

%

%
%
\end{document}